\documentclass[a4paper,11pt]{article}

\usepackage{geometry}										
\usepackage[english]{babel}									
\usepackage[utf8]{inputenc}									
\usepackage[T1]{fontenc}									

\usepackage{lmodern}										

\usepackage{amssymb,amsfonts,amsmath,amsthm}
\usepackage{booktabs}
\usepackage{hyphenat}
\usepackage{booktabs}
\usepackage{bbm}
\usepackage{multirow}
\usepackage{graphicx}
\usepackage{algorithm}
\usepackage{algorithmic}
\usepackage{bstnotations}
\usepackage{xcolor}

\newcommand{\dbar}[1]{\overline{\overline{#1}}}

\newcommand{\I}{\mathbbm{1}_{\cd_f}}

\usepackage{authblk}										

\title{Reliability analysis of non-deterministic models using stochastic emulators}
\author[1]{Anderson V. Pires \thanks{apires@ethz.ch}}
\author[1]{M. Moustapha \thanks{moustapha@ibk.baug.ethz.ch}}
\author[1]{Stefano Marelli\thanks{marelli@ibk.baug.ethz.ch}}
\author[1]{Bruno Sudret\thanks{sudret@ethz.ch}}
\affil[1]{Chair of Risk, Safety and Uncertainty Quantification, ETH Z\"{u}rich, Switzerland}

\date{May 26, 2025}

\usepackage{microtype}										

\usepackage{indentfirst}									
\usepackage{caption}
\usepackage{subcaption}

\usepackage{natbib}
\usepackage{hyperref}
\hypersetup{
pdftitle = {Reliability analysis for data-driven noisy models using active learning},
pdfauthor = {Anderson V. Pires, Maliki Moustapha,  Stefano Marelli, Bruno Sudret},
pdfsubject = {},
pdfkeywords = {Structural reliability, Noisy limit-state functions, Surrogate models, Gaussian process regression, Active learning},
colorlinks = false,                                                                                                                                                                          
}

\usepackage[capitalize]{cleveref}								
\creflabelformat{equation}{#2(#1)#3}								
\crefrangelabelformat{equation}{#3(#1)#4 to #5(#2)#6}						
\labelcrefformat{subequation}{#2(#1)#3}								
\labelcrefrangeformat{subequation}{#3(#1)#4 to #5(#2)#6}					

\graphicspath{{./Figures}} 

\begin{document}

\maketitle

\begin{abstract}
Reliability analysis is a sub-field of uncertainty quantification that assesses the probability of a system performing as intended under various uncertainties. Traditionally, this analysis relies on deterministic models, where experiments are repeatable, \emph{i.e.} they produce consistent outputs for a given set of inputs. However, real-world systems often exhibit stochastic behavior, leading to non-repeatable outcomes. These so-called stochastic simulators produce different outputs each time the model is run, even with fixed inputs.

This paper formally introduces reliability analysis for stochastic models and addresses it by using surrogate models to lower its typically high computational cost. Specifically, we focus on the recently introduced Generalized Lambda Models and Stochastic Polynomial Chaos Expansion. These emulators are designed to learn the inherent randomness of the simulators and enable efficient uncertainty quantification at a much lower cost than traditional Monte Carlo Simulation.

We validate our methodology through three case studies. First, using an analytical function with a closed-form solution, we demonstrate that the emulators converge to the correct solution. Second, we present results obtained from the surrogates using a toy example of a simply supported beam. Finally, we apply the emulators to perform reliability analysis on a realistic wind turbine case study, where only a dataset is available.\\[1em]

{\bf Keywords}: Structural reliability -- Reliability analysis -- Stochastic simulators -- Surrogate models 

\end{abstract}

\maketitle

\newpage

\section{Introduction}
\label{sec_Introduction}

Structural reliability analysis aims to determine the probability that uncertainties in a real-world system lead to performance failure. This analysis typically utilizes computational models that serve as a virtual representation of the actual system. Although these models are often complex and expensive to evaluate, they are widely used because they allow for extensive testing of various scenarios without the need for physical prototypes or experimental setups.

The conditions applied to these virtual models are controlled by a set of input parameters described by a random vector $\ve{X} \in \cd_{\ve{X}} \subset \mathbb{R}^M$, which are characterized by their joint probability density function (PDF) $f_{\ve{X}}$. The state of the system, \emph{i.e.}, whether it fails to meet its performance requirements or not, is determined by the limit-state function $g$, typically associated with the computational model. Failure is assumed when the limit-state function is non-positive, \emph{i.e.}, $g(\ve{x}) \leqslant 0$. By extension, the failure domain is defined as $\cd_f = \acc{\ve{x} : g\prt{\ve{x}} \leqslant 0}$, and the boundary of this domain, where $g\prt{\ve{x}} = 0$, corresponds to the limit-state surface. This framework enables the definition of the probability of failure, $P_f$, as follows \citep{Ditlevsen_1996,Lemaire_2009,Melchers_2018}:
\begin{equation}
	P_f = \Prob{g\prt{\ve{X}} \leqslant 0} = \int_{\cd_f} f_{\ve{X}}(\ve{x}) \, d\ve{x}.
	\label{eq_Pf_deterministic_definition}
\end{equation}

In this context, it is assumed that all uncertainties are captured by the random variables and the limit-state function is deterministic, \emph{i.e.}, evaluating it multiple times with the same inputs yields the same response. However, there are cases where this assumption does not hold.

In contrast to deterministic models, stochastic simulators yield different responses each time they are run with the same input parameters. This variability arises from a source of uncertainty that affects the computational model but cannot be explicitly represented in the input parameters because it is either impractical or undesirable to do so. Common examples of stochastic simulators include epidemiological models, where disease transmission and recovery are treated as random events with specific probabilities \cite{Britton_2010,Zhu_2021_GLaM}. Another example is found in financial modeling, where stock prices are modeled using stochastic processes, such as geometric Brownian motion, to capture market fluctuations and investor behavior \cite{Heston_1993,McNeil_2005,Zhu_2023_SPCE,Luethen_2023}.

As a result of this uncontrollable and inherent randomness, often referred to as \textit{latent variability} or \textit{noise}, stochastic simulators can produce both failed and non-failed outputs for different runs with the same input parameter. In this case, failure does not depend on the input parameter $\ve{x}$ alone but also on the realization of the stochasticity of the model. It is therefore important to account for this latent variability when solving Eq.~\eqref{eq_Pf_deterministic_definition}.

Although research on reliability analysis for problems involving stochastic simulators is in its infancy, there have been a few contributions. \cite{Choe_2015} was among the first to address stochastic simulation within the context of reliability analysis, extending the deterministic importance sampling technique \citep{Melchers_1989} to accommodate stochastic simulators. The authors have continued to expand their work in this field. \cite{Choe_2017} studied the asymptotic properties of the stochastic importance sampling estimator, proposing an asymptotically valid confidence interval. \cite{Cao_2019} introduced a cross-entropy-based solution scheme for stochastic importance sampling. \cite{Pan_2020} introduced an adaptive stochastic importance sampling approach, while \cite{Li_2021} proposed a nonparametric importance sampling method focused on a wind turbine case study. This approach quantifies the contributions of each environmental factor and their interactions while avoiding computational issues with data sparsity in rare event simulation. From another perspective, \cite{Zheng_2022} utilized quantile regression techniques to develop a feedforward neural network surrogate model for stochastic systems. Upon training, they conducted reliability analysis via Monte Carlo simulation (MCS) by sampling across input variables and quantile levels. \cite{Hao_2021} extended the well-known stochastic Kriging method \citep{Ankenman_2010} to handle unknown heteroskedastic Gaussian noise and perform reliability analysis.

In this paper, we formally present structural reliability analysis for problems involving stochastic models. We show that solving this problem via Monte Carlo simulation is possible but, similar to the deterministic case, the approach is too expensive as it requires numerous evaluations of the limit-state function. To mitigate this, we introduce a methodology that employs \emph{stochastic emulators}, \emph{i.e.}, surrogate models for stochastic simulators. These emulators are cheap-to-evaluate models that learn the behavior of stochastic simulators, including their inherent stochasticity, enabling reliability analysis with reduced computational effort. We test the proposed methodology on three case studies. The first is a standard reliability analysis problem converted into a stochastic one by the introduction of latent variables in the limit-state function. The second study applies our approach to a simply supported beam, a classic problem in structural reliability. Analytical solutions are available in both cases, hence allowing us to easily validate our proposed methodology. The third case study focuses on a real-world wind turbine example, where only operational data are available. 

This paper is organized as follows: Section 2 introduces the formal framework for reliability analysis in the context of stochastic simulators. Section 3 presents the stochastic emulators employed in our methodology. Section 4 demonstrates how to perform reliability analysis using these emulators. Section 5 showcases the results of obtained with our methodology when applied to the previously mentioned case studies. Finally, Section 6 presents our conclusions and provides an outlook on future research on the topic.


\section{Problem statement} 
\subsection{Stochastic Simulators}
\label{sec_stochastic_simulators}
A stochastic simulator can be formally represented by the following mapping:
\begin{equation} 
\begin{split} 
g_s : D_{\ve{X}} \times \Omega \quad \to \quad & \mathbb{R}, \\
(\ve{x}, \omega) \quad \mapsto \quad & g_s(\ve{x}, \omega),
\end{split}
\label{eq_map_definition_stochastic_models}
\end{equation}
\noindent where $\omega$ denotes a random event within a probability space $(\Omega, \mathcal{F}, \mathbb{P})$, representing the inherent stochasticity of the system. As discussed in \cite{Luethen_2023}, stochastic simulators can be interpreted from two distinct perspectives.

The first perspective, referred to as the \emph{random variable} point of view, considers that $\ve{x}$ is fixed, causing the response of the simulator to behave as a random variable denoted by $Y_{\ve{x}} = g_s\prt{\ve{x}, \cdot}: \omega \rightarrow \mathbb{R}$. This output can be characterized by its PDF or summarized by quantities such as the mean, variance, and quantiles. The second perspective, referred to as the \emph{random function} point of view, assumes that $\omega$ is fixed, in which case the simulator behaves as a deterministic function $g_s\prt{\cdot, \omega} : D_{\ve{X}} \rightarrow \mathbb{R}$ known as a \emph{trajectory}. Since $\omega$ represents an uncontrollable randomness, this point of view is theoretical and has limited practical applicability.

In practice, modeling true stochasticity is not possible due to the inherently deterministic nature of computers. To mimic the internal randomness of the models, latent variables $\ve{Z}\prt{\omega} \in \mathbb{R}^{M_z}$ are introduced in the simulator. The stochastic simulator becomes then a deterministic function of the pair $\prt{\ve{x},\ve{z}}$, i.e.:
\begin{equation}
Y_{\ve{x}} = g_s\prt{\ve{x},\omega} \equiv g\prt{\ve{x},Z\prt{\omega}}.
\end{equation}
\noindent Each run of the model corresponds to a different random seed, $\omega^{\prt{i}}$, leading to a realization $Y_{\ve{x}}^{\prt{i}}=g\prt{\ve{x},\ve{z}\prt{\omega^{\prt{i}}}}$.

\subsection{Reliability analysis on stochastic simulators}

Because stochasticity is uncontrollable and impacts the outcome of the reliability analysis, it must be accounted for in reliability analysis. This can be addressed directly by incorporating the stochastic simulator into \eqrefe{eq_Pf_deterministic_definition}:
\begin{equation}
P_f = \Prob{g_s\prt{\ve{X}, \omega} \leqslant 0} \equiv \Prob{g\prt{\ve{X},\ve{Z}} \leqslant 0}.
\label{eq_Pf_stochastic_definition}
\end{equation}

The definition in \eqrefe{eq_Pf_stochastic_definition} represents the probability of failure, integrating uncertainty arising from both the input parameters and latent variability inherent to the stochastic simulator. Moreover, equivalent estimators can be formulated based on the two perspectives discussed in Sec.~\ref{sec_stochastic_simulators}. From the random variable point of view, it is possible to define 
the \textit{conditional probability of failure function} $s(\boldsymbol{x})$ as:
\begin{equation}
s(\boldsymbol{x}) = \Probx{\ve{Z}}{g\prt{\ve{X}, \ve{Z} \mid \ve{X}=\ve{x}} \leqslant 0}.
\label{eq_conditional_Pf}
\end{equation}
\noindent This function represents the probability of system failure for a given $\ve{x}$ and solely depends on the intrinsic randomness of the simulator. This formulation is particularly relevant in scenarios where the design is fixed, and the objective is to quantify the impact of the latent stochasticity on the reliability of the system. For example, it could be used to evaluate how different wind excitations affect the reliability of a specific wind turbine.

Considering $s\prt{\ve{x}}$, $P_f$ can be expressed as:
\begin{equation}
P_f = \Espe{\ve{X}}{s(\ve{X})}.
\label{eq_Pf_random_variable_POV}
\end{equation}

On the other hand, the random function perspective leads to:
\begin{equation}
P_f = \Espe{\ve{X}}{\Probx{\ve{Z}}{g\prt{\ve{X}, \ve{Z} \mid \ve{Z}= \ve{z}} \leqslant 0}}.
\label{eq_Pf_random_function_POV}
\end{equation}

\noindent The expression in \eqrefe{eq_Pf_random_function_POV} represents the probability of failure due to uncertainty in the design parameter for a fixed realization of the stochasticity. This approach is particularly useful when stochasticity is kept constant, and the goal is to understand how varying design parameters influence system reliability. For instance, one might investigate how a collection of wind turbines, \emph{i.e.}, a wind farm, responds to the same wind excitation.

\subsection{Estimating $P_f$}

Similarly to the deterministic case, solving \eqrefe{eq_Pf_stochastic_definition} analytically is generally not feasible. An alternative approach is to estimate $P_f$ using Monte Carlo simulation. This method provides an unbiased estimation of $P_f$ because each time the limit-state function is evaluated for a given  $\ve{x}$, a realization of $\omega$ is indirectly obtained. In this sense, MCS is performed on both $\ve{X}$ and $\ve{Z}$, however only $\ve{X}$ is explicitly sampled while $\ve{Z}$ is sampled implicitly through a mechanism internal to the stochastic simulator for each evaluation. Another important aspect is that MCS does not suffer from the curse of dimensionality. Consequently, the presence of stochasticity does not increase the complexity of the problem, and the number of limit-state function evaluations needed only depends on the desired accuracy when estimating $P_f$, just as it does in the deterministic case.

Monte Carlo simulation is widely recognized as often unfeasible, as it requires numerous evaluations of the possibly expensive-to-evaluate limit-state function. Variance-reduction techniques have been developed to address this issue, \emph{e.g.}, importance sampling \citep{Melchers_1989}, subset simulation \citep{Au_2001}, line sampling, or cross-entropy importance sampling \citep{}. These methods aim to provide more accurate estimates of the probability of failure at a lower computational cost compared to crude MCS. While well-established for deterministic simulators, stochastic-aware variance-reduction techniques are not yet widespread, with the only exception being the family of stochastic importance sampling mentioned earlier \citep{Choe_2015}.

Another commonly used approach to reduce the costs associated with MCS is the introduction of surrogate models. These are inexpensive mathematical approximations that mimic the behavior of complex, costly-to-evaluate functions. These models are trained on an \textit{experimental design}, which is a set of points obtained from full-scale simulations. Many surrogate models are available and well-established for deterministic functions. In the context of reliability analysis \citep{Teixeira_2021,Moustapha_2022}, Kriging is likely the most widely used surrogate model. Other methods, such as polynomial chaos expansion \citep{Marelli_2018} can also be employed. However, these approaches cannot be directly applied to stochastic simulators as they do not account for latent stochasticity.

With the growing interest in stochastic simulators, traditional deterministic methods have been extended to handle stochastic cases. \cite{Ankenman_2010} extended the traditional Kriging method to deal with stochastic systems. More recently, \cite{Zhu_2021_GLaM,Zhu_2023_SPCE,Luethen_2023} proposed alternative approaches using polynomial chaos expansions to build surrogate models for stochastic simulators. These recently introduced emulators address various problems and perspectives associated with their respective simulators. In this paper, we capitalize on the methods developed by \cite{Zhu_2021_GLaM,Zhu_2023_SPCE} to perform reliability analysis on stochastic simulators.

\section{Stochastic emulators}


\subsection{Generalized lambda models}
\label{sec_GLaM}
\emph{Generalized lambda models} (GLaM) are a type of surrogate designed for stochastic simulators. The primary objective of GLaM is to mimic the probability density function of the random variable $Y_{\ve{x}} \equiv Y \mid \ve{X} = \ve{x}$. This is done by fitting a \emph{generalized lambda distribution} (GLD) to the conditional response $Y_{\ve{x}}$. GLDs can approximate most common unimodal distributions, such as normal, t-distributed, or Weibull distributions. They are defined by their quantile function $Q\prt{u}$, which is a monotonically increasing function defined on $u \in \bra{0,1}$. GLaM specifically relies on the Freimer-Kollia-Mudholkar-Lin GLD family \citep{Freimer_1988}, which reads:
\begin{equation}
Q(u ; \boldsymbol{\lambda})=\lambda_1+\frac{1}{\lambda_2}\left(\frac{u^{\lambda_3}-1}{\lambda_3}-\frac{(1-u)^{\lambda_4}-1}{\lambda_4}\right),
\label{eq_quantilefunction_GLD}
\end{equation}
where $\lambda_1$ represents the location parameter, $\lambda_2 > 0$ is the scale parameter, $\lambda_3$, and $\lambda_4$ are the shape parameters.  For a deeper explanation of how these parameters affect the resulting GLD, the reader is referred to \cite{Zhu_2020,Zhu_2021_GLaM}.

The rationale behind the GLaM emulator is that for every $\ve{x}$, there exists a set of parameters  $\lambda_{1,\ldots,4}\prt{\ve{x}}$ that can be used to generate a distribution that approximates the response of the original simulator, \emph{i.e.},:
\begin{equation}
Y_{\ve{x}} \sim \operatorname{GLD}\left(\lambda_1(\boldsymbol{x}), \lambda_2(\boldsymbol{x}), \lambda_3(\boldsymbol{x}), \lambda_4(\boldsymbol{x})\right) .
\end{equation}

Each component of $\lambda_{1,\ldots,4}\prt{\ve{x}}$ consists of a deterministic function and, as a consequence, can be modeled by polynomial chaos expansions \citep{Luethen_2021}. Since $\lambda_2\prt{\ve{x}}$ must be positive, its associated expansion is constructed in the logarithm space. The polynomial chaos expansions read:
\begin{equation}
\begin{aligned}
& \lambda_l(\boldsymbol{x}) \approx \lambda_l^{\mathrm{PC}}(\boldsymbol{x} ; \boldsymbol{c})=\sum_{\boldsymbol{\alpha} \in \mathcal{A}_l} c_{l, \boldsymbol{\alpha}} \psi_{\boldsymbol{\alpha}}(\boldsymbol{x}), \quad l=1,3,4, \\
& \lambda_2(\boldsymbol{x}) \approx \lambda_2^{\mathrm{PC}}(\boldsymbol{x} ; \boldsymbol{c})=\exp \left(\sum_{\boldsymbol{\alpha} \in \mathcal{A}_2} c_{2, \boldsymbol{\alpha}} \psi_{\boldsymbol{\alpha}}(\boldsymbol{x})\right),
\end{aligned}
\end{equation}
\noindent where $\psi_{\ve{\alpha}}\prt{\ve{x}}$ are a set of polynomials, orthonormal with respect to the input distribution $f_{\ve{X}}$, $\ve{\alpha}$ are multi-indices, which identify the degree of the multivariate polynomials, $c = \{c_{l,\alpha}: l = 1, \ldots, 4, \alpha \in \mathcal{A}_l\}$ denote the corresponding coefficients and $\mathcal{A} = \{\mathcal{A}_l: l = 1, \ldots, 4\}$ represents the truncation set, which defines the basis functions. For a more detailed explanation on the construction of PCEs and their properties, interested readers are referred to \cite{Luethen_2021}.

There are two approaches to constructing the GLaM emulator. The first approach, introduced by \cite{Zhu_2020}, relies on replications, \emph{i.e.}, for each point $\ve{x}^{\prt{i}}$, there are $R$ evaluations of the simulator: $ \mathcal{Y}^{(i)} = \acc{y^{(i,1)},y^{(i,2)},\ldots,y^{(i,R)}}$. The training involves then two steps. First, the local GLD parameters $\lambda_{1,\ldots,4}\prt{\ve{x}^{(i)}}$ for each $\ve{x}^{(i)}$  are inferred from the dataset $\mathcal{Y}^{(i)}$, using either the method of moments or maximum likelihood estimation. In the second step, PCE approximation of the lambda functions, \emph{i.e.} the mapping $\ve{x} \mapsto \lambda_{1,\ldots,4}\prt{\ve{x}}$, are constructed using the experimental design $\acc{\prt{\ve{x}^{(i)},\,\lambda_{1,\ldots,4}\prt{\ve{x}^{(i)}}}, \, i = 1,\ldots, N}$.

A second, generally more efficient approach was introduced by \cite{Zhu_2021_GLaM} and eliminates the need for replications. Given an experimental design $\cx = \acc{\ve{x}^{\prt{1}}. \ldots, \ve{x}^{\prt{N}}}$ and its associated model responses $\cy = \acc{y^{\prt{1}}. \ldots,y^{\prt{N}}}$, the associated maximum likelihood optimization problem is directly cast as a function of the PCE coefficients $\ve{c}$:
\begin{equation}
  \hat{c}=\arg \max _{\boldsymbol{c} \in \cc} \mathrm{L}(\boldsymbol{c}).
\end{equation}
\noindent where
\begin{equation}
    \mathrm{L}(\boldsymbol{c})=\sum_{i=1}^N \log \left(f\left(y^{(i)} ; \boldsymbol{\lambda}^{\mathrm{PC}}\left(\boldsymbol{x}^{(i)} ; \boldsymbol{c}\right)\right)\right).
    \label{eq_estimator_KL_div}
\end{equation}

\noindent In \eqrefe{eq_estimator_KL_div}, $f$ represents the PDF of the generalized lambda distribution, which can be obtained by numerically calculating the reciprocal of the derivative of the quantile function in \eqrefe{eq_quantilefunction_GLD}. The set $\cc$ defines the search space for $\ve{c}$. For further details on the estimator in \eqrefe{eq_estimator_KL_div}, refer to \cite{Zhu_2020,Zhu_2021_GLaM}.


A detailed analysis of the impact of replications on experimental design is provided in \cite{Zhu_2021_GLaM}. A straightforward advantage of not using replications is that the available computational budget can be allocated to a more space-filling experimental design, ensuring information is distributed across the entire domain rather than concentrated at replicated points. This leads to a more comprehensive and robust model. Furthermore, it eliminates the need for predefined experimental design constraints, allowing datasets collected for other purposes to be used to train the emulator.

\subsection{Stochastic polynomial chaos expansion}
\label{sec_SPCE}

\emph{Stochastic Polynomial Chaos Expansion} (SPCE), introduced by \cite{Zhu_2023_SPCE}, is another stochastic emulator that enables modeling the output distribution of $Y_{\ve{x}}$. It works by introducing an artificial latent variable $Z$ to represent the stochasticity of the simulator within the polynomial chaos expansion framework. This approach capitalizes on the Probability Integral Transform (PIT), a statistical technique that allows transforming one continuous random variable into another, enabling the mapping of a known variable $Z$ to the desired output distribution. Typically, the choice for the distribution of $Z$ is between a standard uniform and a standard Gaussian.

Let the cumulative distribution function (CDF) associated to $Y_{\ve{x}}$ be denoted by $F_{Y \mid \boldsymbol{X}}(y \mid \boldsymbol{x})$. The PIT allows us to express a relationship between $Y_{\ve{x}}$ and an artificial latent variable $Z$, as follows:
. The symbol $\stackrel{\mathrm{d}}{=}$ indicates that the two sides are equal in distribution.

\eqrefe{eq_PIT} implies that $Y_{\boldsymbol{x}}$ can be represented as a deterministic function of both the input $\boldsymbol{x}$ and the latent variable $Z$. This deterministic mapping is then captured using a PCE in the $\prt{\boldsymbol{X}, Z}$ space:
\begin{equation}
F_{Y \mid \boldsymbol{X}}^{-1}\left(F_Z(Z) \mid \boldsymbol{X}\right) \stackrel{\mathrm{d}}{=} \sum_{\boldsymbol{\alpha} \in \mathbb{N}^{M+1}} c_{\boldsymbol{\alpha}} \psi_{\boldsymbol{\alpha}}(\boldsymbol{X}, Z).
\label{eq_PIT_SPCE}
\end{equation}

For a given \( \boldsymbol{x} \), the expansion above is a function of the latent variable \( Z \), making the response \( Y_{\boldsymbol{x}} \) a random variable. Considering a truncation scheme \( \mathcal{A} \), we have:
\begin{equation}
Y_x \stackrel{\mathrm{d}}{\approx} \tilde{Y}_x=\sum_{\alpha \in \mathcal{A}} c_\alpha \psi_{\boldsymbol{\alpha}}(\boldsymbol{x}, Z).
\label{eq_ill_posed_SPCE}
\end{equation}

Relying on \eqrefe{eq_ill_posed_SPCE} as-is may lead to issues such as singularities in the probability density functions. To address this, \cite{Zhu_2023_SPCE} introduced an additive noise component $\epsilon \sim \mathcal{N}(0, \sigma^2)$, which acts, in practice, as a regularization term. The resulting emulator is then expressed as:
\begin{equation}
Y_x \stackrel{\mathrm{d}}{\approx} \tilde{Y}_x=\sum_{\alpha \in \mathcal{A}} c_\alpha \psi_\alpha(\boldsymbol{x}, Z)+\epsilon.
\end{equation}

The PDF of $\tilde{Y}_{\boldsymbol{x}}$ results from the convolution of the PCE with the Gaussian noise. Assuming the noise variance $\sigma^2$ is known, the corresponding PDF is:
\begin{equation}
f_{\tilde{Y}_{\boldsymbol{x}}}(y) =\int_{\mathcal{D}_Z} \frac{1}{\sigma} \varphi\left(\frac{y-\sum_{\boldsymbol{\alpha} \in \mathcal{A}} c_{\boldsymbol{\alpha}} \psi_{\boldsymbol{\alpha}}(\boldsymbol{x}, z)}{\sigma}\right) f_Z(z) \mathrm{d} z,
\label{eq_PDF_SPCE}
\end{equation}

\noindent where $\varphi$ represents the standard normal PDF.

Given a replication-free experimental design, as defined in Sec.~\ref{sec_GLaM}, the PCE coefficients $\ve{c}$ can be computed using maximum likelihood estimation. Because the noise variance is, in general, not known in advance, tuning this parameter is still needed. \cite{Zhu_2023_SPCE} proposes using cross-validation for this task. A more detailed explanation regarding the fitting process can be found in \cite{Zhu_2023_SPCE}.

SPCE does not require replications in the experimental design, making all the advantages discussed in Sec.~\ref{sec_GLaM} also applicable. Moreover, unlike GLaM, SPCE does not impose a specific parametric form on the simulator response. Instead, it offers a flexible model capable of approximating a wide range of distributions, including bi-modal ones. Another notable advantage of SPCE is that it represents stochasticity using a single latent variable. Regardless of the complexity or the number of latent variables needed to capture the inherent stochasticity of the model, the emulator consistently reduces it to a single latent variable.

\section{Proposed methodology for reliability analysis}
\label{sec_reliability_on_emulators}
Once the emulators are trained, the probability of failure $P_f$ can be estimated using MCS. Since $P_f$ can be defined in various ways, as in \cref{eq_Pf_stochastic_definition,eq_Pf_random_variable_POV,eq_Pf_random_function_POV}, we need to determine which definition yields the estimator with the lowest variance. To achieve this, we compare the variances of the MCS estimators derived from \cref{eq_Pf_stochastic_definition,eq_Pf_random_variable_POV}. This comparison is only feasible because $\hat{s}(\ve{x})$, the emulator-based approximation of $s(\ve{x})$, is both efficient to compute and numerically precise. As a related consideration, the accuracy of $\hat{s}(\ve{x})$ is strongly dependent on the quality of the emulators.

Estimating \eqrefe{eq_Pf_stochastic_definition} follows the typical MCS procedure and can be done from both the simulator or the emulator. We introduce an indicator function $\I(\ve{x}, \ve{z})$, which takes the value $1$ for a failed outcome and $0$ otherwise. Next, we generate $N_{\text{MCS}}$ random samples $\cx_{\text{MCS}} = \acc{\boldsymbol{x}^{(1)}, \dots, \boldsymbol{x}^{N_{\text{MCS}}}}$ according to $f_{\ve{X}}$, run the model for each sample, and evaluate the associated indicator function.

$\overline{P_f}$ denotes the MCS estimator of \eqrefe{eq_Pf_stochastic_definition}. It represents the probability of failure associated with the considered model and corresponds to the empirical mean of the indicator function. The variance of this estimator is given by:
\begin{equation}
\begin{split}
\Var{\overline{P_f}} &= \Var{\frac{1}{N_{\text{MCS}}} \sum_{i=1}^{N_{\text{MCS}}} \I(\boldsymbol{x}_i, \boldsymbol{z}_i)},\\
&= \frac{1}{N_{\text{MCS}}} \Var{\I(\boldsymbol{X}, \boldsymbol{Z})}.
\label{var_single_loop}
\end{split}
\end{equation}

Similarly, for the definition in \eqrefe{eq_Pf_random_variable_POV}, the Monte Carlo estimator $\hat{P}_f$ has the associated variance:
\begin{equation}
\begin{split}
\Var{\hat{P_f}} &= \Var{\frac{1}{N_{\text{MCS}}} \sum_{i=1}^{N_{\text{MCS}}} \hat{s}(\boldsymbol{x}_i)},\\
&= \frac{1}{N_{\text{MCS}}} \Var{\hat{s}(\boldsymbol{X})}.
\end{split}
\end{equation}

Thus, for a fixed $N_{\text{MCS}}$, comparing the variance of the estimators reduces to comparing $\Var{\I(\boldsymbol{X}, \boldsymbol{Z})}$ and $\Var{\hat{s}(\boldsymbol{X})}$. From the law of total variance, we have:
\begin{equation}
\begin{split}
\Var{\I(\mathbf{X}, \mathbf{Z})} &= \mathbb{E}_{\mathbf{X}}\left[\operatorname{Var}_{\mathbf{Z}}(\I(\mathbf{X}, \mathbf{Z}) \mid \mathbf{X})\right] + \operatorname{Var}_{\mathbf{X}}\left(\mathbb{E}_{\mathbf{Z}}[\I(\mathbf{X}, \mathbf{Z}) \mid \mathbf{X}]\right),\\
&= \mathbb{E}_{\mathbf{X}}\left[\operatorname{Var}_{\mathbf{Z}}(\I(\mathbf{X}, \mathbf{Z}) \mid \mathbf{X})\right] + \operatorname{Var}_{\mathbf{X}}(\hat{s}(\mathbf{X})).
\end{split}
\end{equation}

This implies that, for a fixed computational budget, $\Var{\overline{P}_f}$ is necessarily larger than $\Var{\hat{P_f}}$, as the first term on the right-hand side is always non-negative. This reduction in variance arises because the emulators enable numerically integrating the uncertainty associated with the latent space $\ve{Z}$. However, it is important to note that this advantage only exists when $s(\ve{x})$ can be efficiently and precisely computed, as in the case of SPCE and GLaM emulators. We show in Annex~\ref{annex_A} that, in the general case where $s\prt{\ve{x}}$ is unknown or difficult to compute, the estimator $\overline{P_f}$ is preferable.

\subsection{Computing $\hat{s}(\ve{x})$ from GLaM emulator}

For the generalized lambda models, computing the conditional probability of failure $\hat{s}(\ve{x})$ is straightforward. Once trained, the PCE model provides the parameters $\lambda_{1,\ldots,4}(\ve{x})$, and the associated quantile function of the generalized lambda distribution is obtained according to \eqrefe{eq_quantilefunction_GLD}. The computation of $\hat{s}(\ve{x})$ is then given by:
\begin{equation}
    \hat{s}\prt{\ve{x}} = Q^{-1}\prt{0; \lambda_{1,\ldots,4}\prt{\ve{x}}}.
\end{equation}

Since the CDF of the generalized lambda distribution does not have a closed-form solution, a numerical inversion of the quantile function is required. Due to the monotonic property of the quantile function, this inversion is computationally efficient.

\subsection{Computing $\hat{s}\prt{\ve{x}}$ from SPCE emulator}

For stochastic polynomial chaos expansion, the CDF associated with a predicted $Y_{\ve{x}}$, denoted as $F_{\tilde{Y}_{\ve{x}}}\prt{y}$, is obtained by integrating the PDF introduced in \eqrefe{eq_PDF_SPCE}. Since this PDF is a one-dimensional integral, it can be evaluated using Gaussian quadrature as follows:
\begin{equation}
\begin{split}
f_{\tilde{Y}_{\boldsymbol{x}}}(y) &=\int_{\mathcal{D}_Z} \frac{1}{\sigma} \varphi\left(\frac{y-\sum_{\boldsymbol{\alpha} \in \mathcal{A}} c_{\boldsymbol{\alpha}} \psi_{\boldsymbol{\alpha}}(\boldsymbol{x}, z)}{\sigma}\right) f_Z(z) \mathrm{d} z\\
&\approx \sum_{j=1}^{N_Q} \frac{1}{\sqrt{2 \pi} \sigma} \exp \left(-\frac{\left(y-\sum_{\boldsymbol{\alpha} \in \mathcal{A}} c_{\boldsymbol{\alpha}} \psi_{\boldsymbol{\alpha}}\left(\boldsymbol{x}, z_j\right)\right)^2}{2 \sigma^2}\right) w_j,
\end{split}
\end{equation}
\noindent where $N_Q$ is the number of integration points, $z_j$ is the $j$-th integration point, and $w_j$ is the corresponding weight, both associated with the weight function $f_Z$.

The approximation introduced by the numerical integration scheme implies that $f_{\tilde{Y}_{\boldsymbol{x}}}(y)$ is a mixture of $N_Q$ Gaussian distributions $\acc{\mathcal{N}\left(\sum_{\boldsymbol{\alpha} \in \mathcal{A}} c_{\boldsymbol{\alpha}} \psi_{\boldsymbol{\alpha}}\left(\boldsymbol{x}, z_j\right), \sigma^2 \right), \, j = 1 \ldots N_Q}$, with corresponding weights $w_j$. The CDF of this approximation is straightforward and reads:
\begin{equation}
\begin{split}
F_{Y_{\ve{x}}}(y) &\approx \int_{- \infty}^y \sum_{j=1}^{N_Q} \frac{1}{\sqrt{2 \pi} \sigma} \exp \left(-\frac{\left(y-\sum_{\boldsymbol{\alpha} \in \mathcal{A}} c_{\boldsymbol{\alpha}} \psi_{\boldsymbol{\alpha}}\left(\boldsymbol{x}, z_j\right)\right)^2}{2 \sigma^2}\right) w_j dy, \\
&\approx \sum_{j=1}^{N_Q} w_j  \int_{- \infty}^y  \frac{1}{\sqrt{2 \pi} \sigma} \exp \left(-\frac{\left(y-\sum_{\boldsymbol{\alpha} \in \mathcal{A}} c_{\boldsymbol{\alpha}} \psi_{\boldsymbol{\alpha}}\left(\boldsymbol{x}, z_j\right)\right)^2}{2 \sigma^2}\right) dy, \\
&\approx  \sum_{j=1}^{N_Q} w_j \Phi\prt{\frac{y-\sum_{\boldsymbol{\alpha} \in \mathcal{A}} c_{\boldsymbol{\alpha}} \psi_{\boldsymbol{\alpha}}\left(\boldsymbol{x}, z_j\right)}{\sigma}},
\end{split}
\end{equation}
where $\Phi$ is the standard Gaussian CDF.

Finally, the conditional probability of failure function is given by:
\begin{equation}
    \hat{s}\prt{\ve{x}} \approx  \sum_{j=1}^{N_Q} w_j \Phi\prt{-\frac{\sum_{\boldsymbol{\alpha} \in \mathcal{A}} c_{\boldsymbol{\alpha}} \psi_{\boldsymbol{\alpha}}\left(\boldsymbol{x}, z_j\right)}{\sigma}}. 
\end{equation}

Once the SPCE model is trained, computing $\sum_{\boldsymbol{\alpha} \in \mathcal{A}} c_{\boldsymbol{\alpha}} \psi_{\boldsymbol{\alpha}}(\boldsymbol{x}, z)$ is inexpensive. The accuracy of this approach strongly depends on the number of quadrature points used. In this contribution, we used $N_Q = 100$, which, based on our tests, results in a negligible numerical error.

\section{Results}
All simulations presented in this section were conducted using the Generalized Lambda Models and Stochastic Polynomial Chaos Expansions modules from the UQLab software for uncertainty quantification \citep{UQdoc_21_120,UQdoc_21_121,Marelli_2014_UQLab}.

\subsection{Stochastic $R-S$ function}
We first showcase the usage of the presented stochastic emulators using a modified stochastic $R-S$ example. The limit-state function reads:
\begin{equation}
g(\ve{X}, \ve{Z}) = \frac{R}{Z_1} - S\cdot Z_2,
\label{eq_stochastic_RS}
\end{equation}
where $\ve{X} = \acc{R,S}$ are the random variables representing the resistance and demand, as in the classical $R-S$ problem, and $\ve{Z} = \acc{Z_1,Z_2}$ are two latent variables introduced to represent the inner stochasticity of the model. The distribution of the associated random variables, their moments and their associated parameters are described in Table~\ref{tab_RS}.
\begin{table}[H]
\centering
\caption{Moments, distributions, and parameters of the considered variables for the $R-S$ problem.}
\label{tab_RS}
\begin{tabular}{@{}cccccc@{}}
\toprule
Variable & Distribution & Mean & Std. Deviation & $\lambda$              & $\zeta$  \\ \midrule
$R$      & Lognormal    & 5.0    & 0.8            & 1.5968                 & 0.1590   \\
$S$      & Lognormal    & 2.0    & 0.6            & 0.6501                 & 0.2936   \\
$Z_1$    & Lognormal    & 1.0    & 0.028          & -0.0004 & 0.0280   \\
$Z_2$    & Lognormal    & 1.0    & 0.096          & -0.0046                & 0.0958   \\ \bottomrule
\end{tabular}
\end{table}

Given the problem setup, an analytical solution can be obtained by applying a logarithmic transform to each component of the equation. This yields the following analytical solution:
\begin{equation}
P_f = \Phi \left(\frac{ \lambda_{R} - \lambda_{Z_1} - \lambda_{S} -\lambda_{Z_2} }{\sqrt{\zeta_{R}^2 + \zeta_{Z_1}^2 + \zeta_{S}^2 + \zeta_{Z_2}^2}}\right) = 3.154 \times 10^{-3} .
\label{eq_true_Pf_RS}
\end{equation}

It is also possible to obtain an analytical solution for the conditional probability of failure function $s\prt{\ve{x}}$, defined in \eqrefe{eq_conditional_Pf}. In this case, we condition the limit-state function on $\ve{X}$, while $\ve{Z}$ is random. This leads to the following function:
\begin{equation}
s\prt{\ve{x}} = s\prt{r,s} = \Phi \prt{\frac{\ln{r} -\lambda_{Z_1} -\ln{s} -\lambda_{Z_2}}{\sqrt{\zeta_{Z_1}^2 + \zeta_{Z_2}^2}}}.
\label{eq_sX_RS}
\end{equation}

To demonstrate the performance of the emulators in estimating $P_f$ and $s(\ve{x})$, we trained both models using various experimental design sizes, sampled via space-filling Latin Hypercube Sampling \citep[LHS,][]{Olsson_2003}, with sizes $N \in \{500; 1{,}000; 5{,}000; 10{,}000; 50{,}000\}$. The polynomial chaos expansions used in surrogating the parameters $\lambda_{1,2}\prt{\ve{x}}$ of the GLaM model have degree adaptivity ranging from $p \in \bra{0, 3}$, while the expansions related to parameters $\lambda_{3,4}\prt{\ve{x}}$ vary from $p \in \bra{0, 2}$. The SPCE emulators have degree adaptivity ranging from $p \in \bra{0, 4}$. For both emulators, the hyperbolic truncation q-norm ranges in $q \in \bra{0.7,1}$ with constant increments of $0.1$. 

The probability of failure $\hat{P}_f$ is estimated for each emulator using Monte Carlo simulation with a fixed sample size of $N_{\text{MCS}} = 10^6$. Moreover, to compare the performance of the emulators against direct MCS, we also estimate $\overline{P_f}$ using the same sample size of full-scale simulations that were used to train the emulators, \emph{i.e.} $N \in \{500; 1{,}000; 5{,}000; 10{,}000; 50{,}000\}$. This allows us to demonstrate that, with the same number of full-scale simulations, the emulators produce lower variance in their estimations. We depict the results in box plots, as each experimental setup is repeated 50 times to capture statistical variability. The width of the box plots represents the interquartile range, the horizontal line is the median and responses deviating by more than $\pm 2.7$ standard deviations from the mean are considered outliers and plotted as crosses.

\begin{figure}[H]
     \centering
     \begin{subfigure}[c]{0.49\textwidth}
         \centering
         \includegraphics[width=0.9\textwidth]{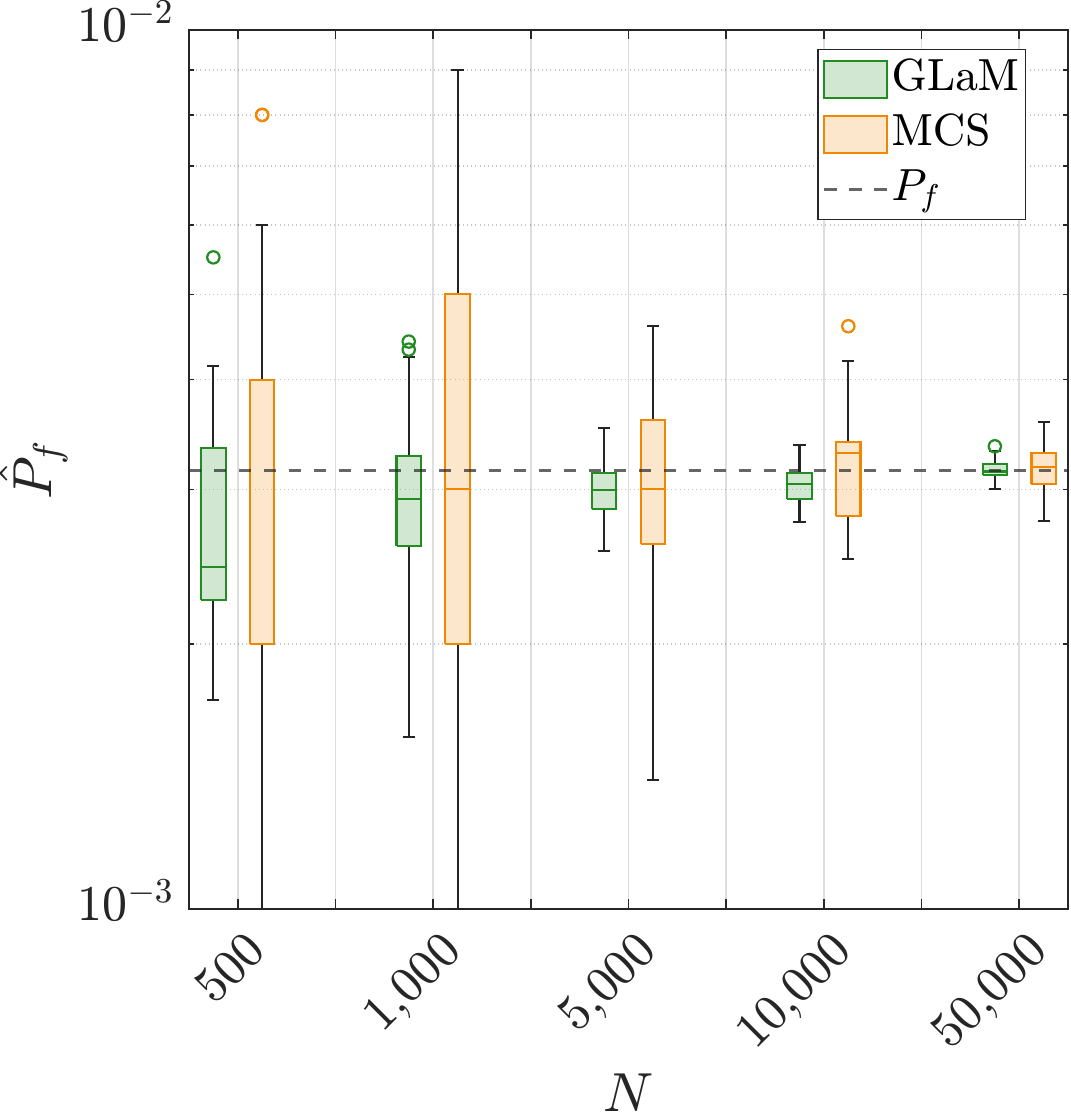}
         \caption{}
    \label{fig_RS_Pf_GLaM}
     \end{subfigure}
     \begin{subfigure}[c]{0.49\textwidth}
         \centering
         \includegraphics[width=0.9\textwidth]{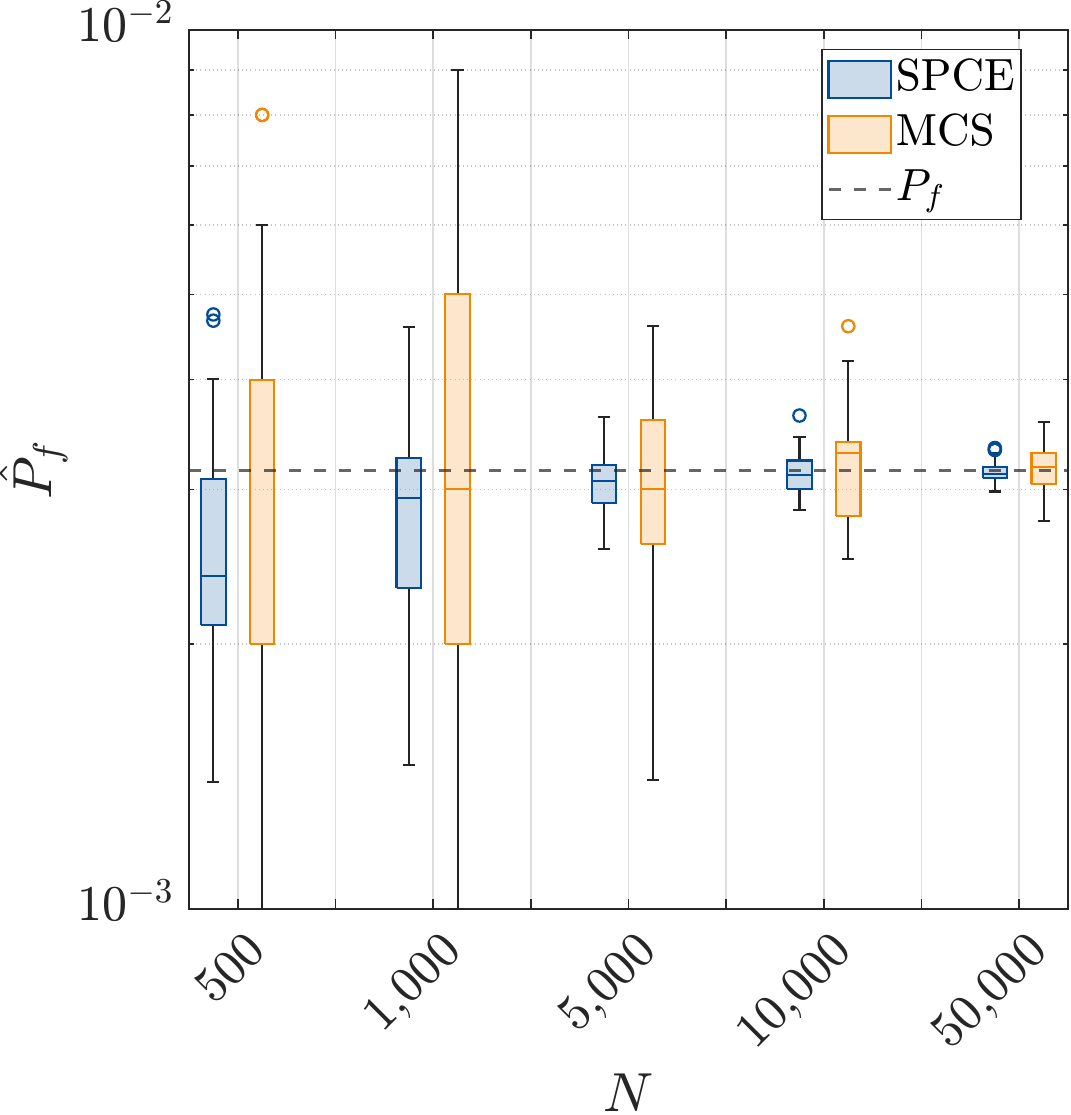}
         \caption{}
    \label{fig_RS_Pf_SPCE}
     \end{subfigure}
     \caption{$R-S$ function -- (a): Box plots comparing convergence behavior obtained from GLaM emulator (in green) and from direct MCS (in orange) for increasing values of $N$. The analytical probability of failure is depicted by the dashed black line. (b): Box plots comparing convergence behavior obtained from SPCE emulator (in blue) and from direct MCS (in orange) for increasing values of $N$. The analytical probability of failure is depicted by the dashed black line.}
	 \label{fig_Pf_RS}
\end{figure}

\figref{fig_Pf_RS} depicts the box plots of the probabilities of failure estimated by the emulators, with GLaM in green and SPCE in blue, alongside the MCS estimates in orange. The black dashed line represents the true probability of failure $P_f$ (\eqrefe{eq_true_Pf_RS}). The emulators provide relatively accurate estimates of $P_f$, even in a small data regime when $N \leqslant 1{,}000$ samples. As the size of the experimental design increases, the bias on the estimations of the emulators disappears and the emulators converge to $P_f$. Additionally, the box plots resulting from the emulators are consistently narrower than those from direct MCS, demonstrating the variance-reduction property of the emulators. These results suggest that the emulators are a viable option for performing reliability analysis.

The current implementation of this methodology still requires numerous evaluations of the limit-state function. However, active learning approaches, as shown in the literature for deterministic models \citep{Teixeira_2021, Moustapha_2022}, offer a way to significantly reduce the cost of training emulators. By iteratively enriching the experimental design on regions where a more accurate surrogate model is needed, these methods can drastically lower the number of simulations required. As a result, precise estimates of $P_f$ can be obtained with relatively small experimental designs, greatly minimizing computational costs. To the author's best knowledge, such an approach is still not available in the context of structural reliability for stochastic emulators.

\begin{figure}[H]
         \centering
         \includegraphics[width=\textwidth]{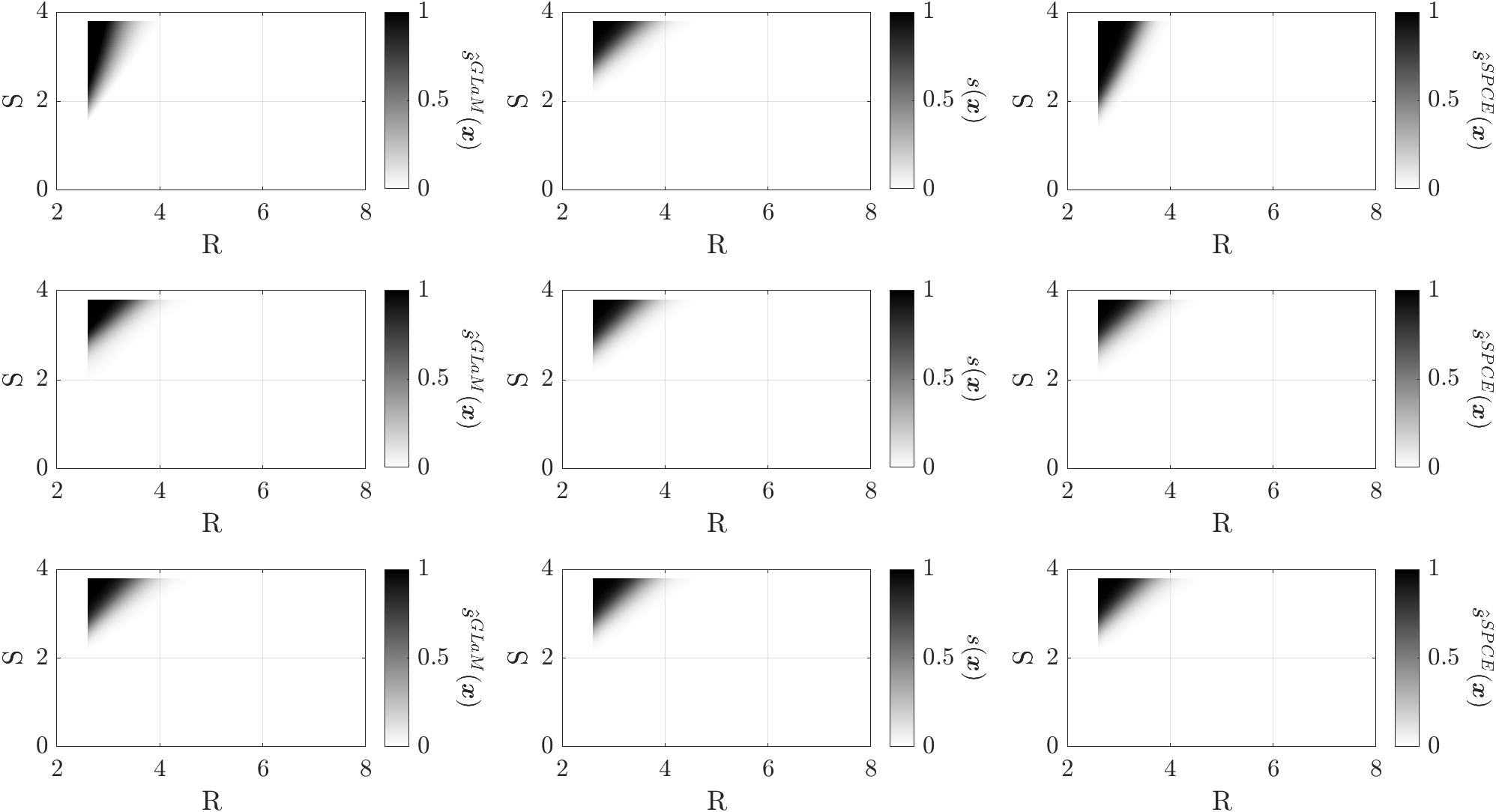}
         \caption{$R-S$ function: Comparison of conditional probability of failure function $s\prt{\ve{x}}$ across different estimation methods and sample sizes. Each plot depicts the heatmap of $s\prt{\ve{x}}$ with a different method and sample sizes. The columns represent different estimation methods: the first column displays results from GLaM, the second column shows the reference values obtained analytically, and the third column illustrates results from SPCE. The rows correspond to different training sample sizes used for the emulators: the first row uses $500$ points, the second row uses $5{,}000$ points, and the third row uses $50{,}000$ points.}
    \label{fig_RS_cond_Pf}
\end{figure}

Due to the stochastic nature of the problem, depicting a limit-state surface is not possible. However, to visually compare the performance of the emulators, we plot heat maps of the function $s(\ve{x})$ and those estimated by GLaM and SPCE, denoted as $s^{GLaM}(\ve{x})$ and $s^{SPCE}(\ve{x})$, respectively. \figref{fig_RS_cond_Pf} displays the heat maps for three different experimental design sizes: $N = \acc{500; 5{,}000; 50{,}000}$. These surfaces correspond to the seeds that led to the median $\hat{P}_f$ in \figref{fig_Pf_RS}. For $N = 500$, there is a noticeable difference between the reference function and those produced by the emulators. However, as the size of the training set increases, the emulators become more accurate, with only minor differences observed for $N = 5{,}000$. For $N = 50{,}000$, no distinguishable differences are evident.

\subsection{Stochastic simply supported beam}
Let us consider a simply supported beam subjected to uniform load. We are interested in the probability that the mid-span deflection exceeds a given threshold $t_{lim}$. The associated limit-state function reads:
\begin{equation}
g(\ve{X}, E) = t_{lim} - \prt{\frac{5pL^4}{32Ebh^3}}.
\label{eq_SSB_limit_state}
\end{equation}
The input variables are $\ve{X} = \{p, L, b, h\}$, and $Z = E$ is treated as latent variable. By doing this, we aim to simulate a scenario where heterogeneous properties of the materials are modeled using random fields \citep{Grigoriu_2002}, which would naturally require a stochastic treatment. However, since we focus on validating the proposed methodology in an academic context, we do not discuss appropriate ways of modeling these properties. Instead, we simplify the problem by using $E$ as a latent variable. The distributions of the related random variables, along with their moments and corresponding parameters, are presented in Table~\ref{tab_SSB}.
\begin{table}[H]
  \centering
  \caption{Moments, distributions, and parameters of the considered variables for the simply supported beam example.}
  \label{tab_SSB}
  \begin{tabular}{@{}cccccc@{}}
\toprule
Variable & Distribution & Mean & Std. Deviation & $\lambda$              & $\zeta$  \\ \midrule
$b$      & Lognormal    & 0.15 m    & $7.5\times 10^{-3}$ m            & -1.8984                 & 0.0500   \\
$h$      & Lognormal    & 0.3 m    & $15\times 10^{-3}$ m            & -1.2052                 & 0.0500  \\
$L$    & Lognormal    & 5.0 m   & 0.05 m          & 1.6094 & 0.0100   \\
$p$    & Lognormal    & 10 kN/m    & 2 kN/m          & 9.1907                & 0.1980   \\
$E$    & Lognormal    & $30{,}000$ MPa   & $4{,}500$ MPa          & 24.1133                & 0.1492   \\  \bottomrule
\end{tabular}
\end{table}
The probability of failure can be calculated analytically and reads: 
\begin{equation}
P_f= \Phi\prt{\frac{\ln(t_{lim}) - \ln\left(\frac{5}{32}\right) - \lambda_p - 4\lambda_L + \lambda_E + \lambda_b + 3\lambda_h}{\sqrt{\zeta_p^2 + 16\zeta_L^2 + \zeta_E^2 + \zeta_b^2 + 9\zeta_h^2}}}.
\end{equation}
We consider a displacement threshold of $t_{lim}= \textrm{span}/{250} = 0.02$ m, according to \cite{standard2004eurocode}. This yields \mbox{$P_f=1.019 \times 10^{-3}$}.

We evaluate the performance of the emulators using a similar approach as in the previous example. Both emulators were trained with different experimental design sizes $N \in \acc{500; 5{,}000; 10{,}000; 50{,}000}$ obtained via LHS \citep{Olsson_2003}. The polynomial chaos expansions used to approximate the parameters $\lambda_{1,2}(\ve{x})$ in the GLaM model had adaptive degrees ranging from $p \in [1, 4]$, while the expansions for $\lambda_{3,4}(\ve{x})$ varied from $p \in [0, 2]$. For SPCE, the degree adaptivity ranged from $p \in [1, 7]$. The hyperbolic truncation q-norm for all expansions ranged from $q \in [0.7, 1]$, with increments of $0.1$. $\hat{P}_f$ was computed via MCS with $N_{MCS} = 10^6$ samples. To compare with direct MCS, we also estimated $\overline{P_f}$ using the same number of full-scale simulations for training the emulators. In practice, this corresponds to using the experimental design to estimate $\overline{P_f}$. All results are displayed in box plots as we carried out 50 experiment repetitions to account for the statistical variability.  

\begin{figure}[H]
     \centering
     \begin{subfigure}[c]{0.49\textwidth}
         \centering
         \includegraphics[width=0.9\textwidth]{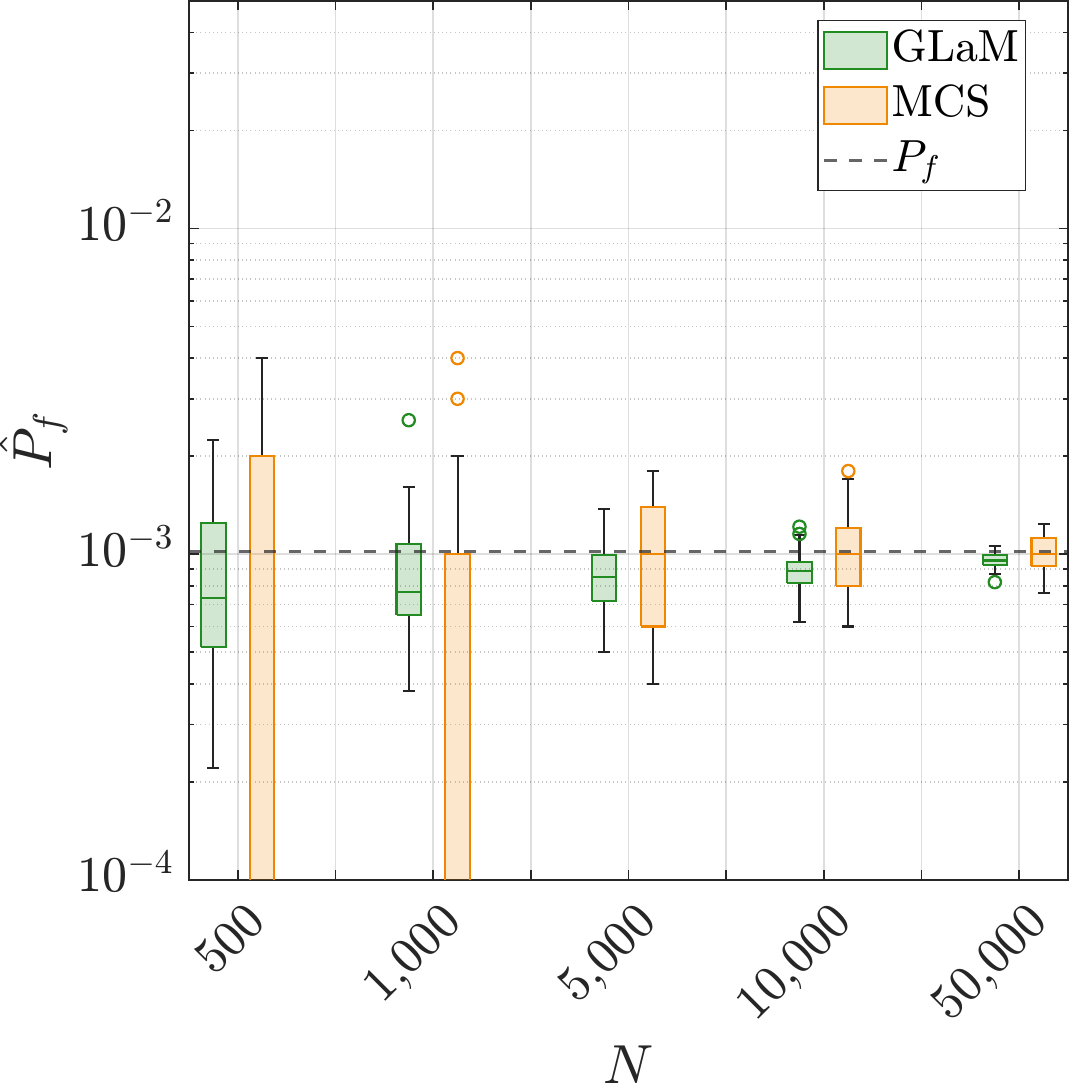}
         \caption{}
    \label{fig_SSB_Pf_GLaM}
     \end{subfigure}
     \begin{subfigure}[c]{0.49\textwidth}
         \centering
         \includegraphics[width=0.9\textwidth]{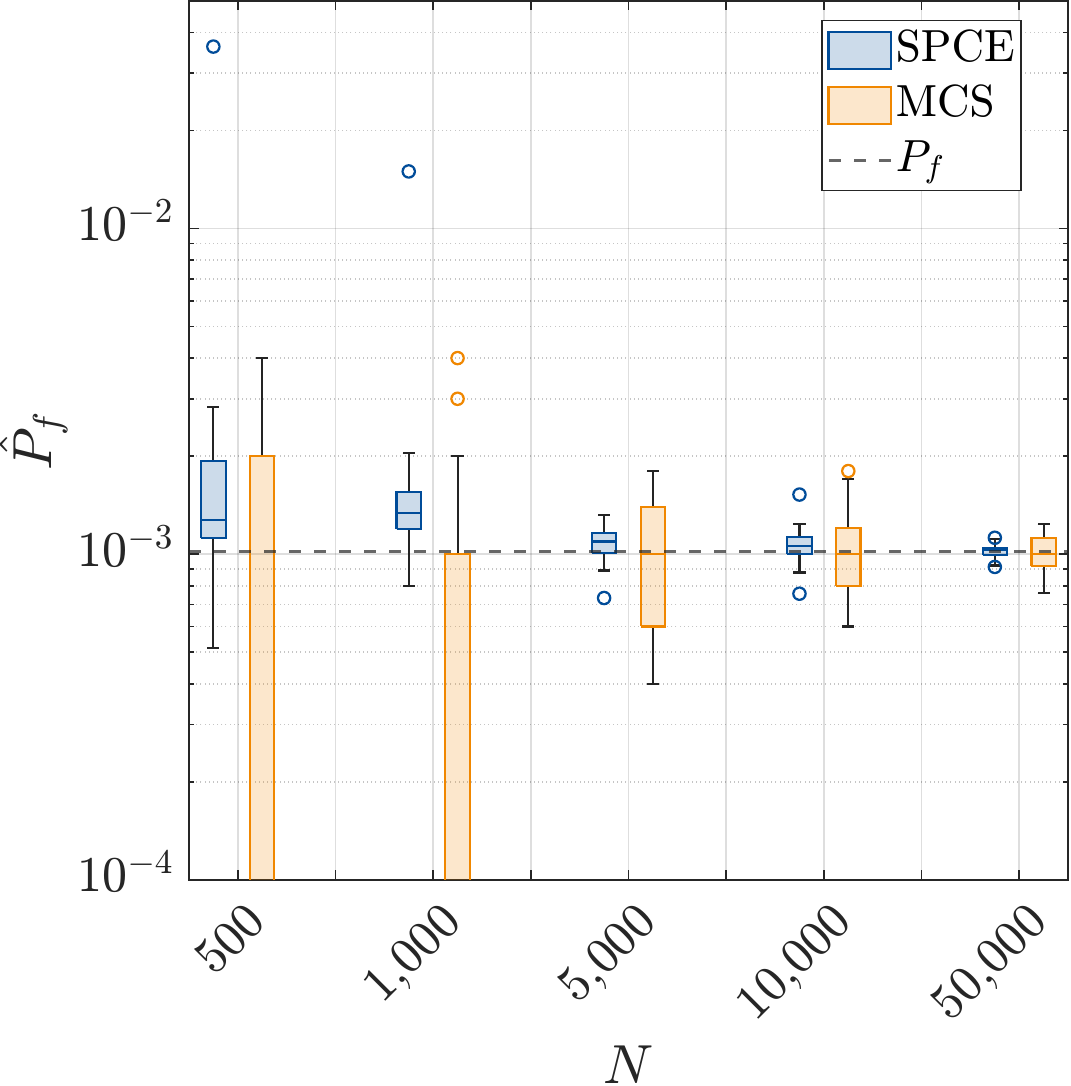}
         \caption{}
    \label{fig_SSB_Pf_SPCE}
     \end{subfigure}
     \caption{Simply supported beam example -- (a): Box plots comparing convergence behavior obtained from GLaM emulator (in green) and from direct MCS (in yellow) for increasing values of $N$. The analytical probability of failure is depicted by the dotted line. (b): Box plots comparing convergence behavior obtained from SPCE emulator (in blue) and from direct MCS (in yellow) for increasing values of $N$. The analytical probability of failure is depicted by the dotted line.}
	 \label{fig_Pf_SSB}
\end{figure}

\figref{fig_Pf_SSB} presents the results from the simulations. The emulators provide relatively accurate estimates of $P_f$, even with small data sets ($N \leqslant 1{,}000$), despite a slight bias. In contrast, for many Monte Carlo simulation runs with $N \leqslant 1{,}000$, no failed samples were observed, leading to $\overline{P}_f = 0$, and producing the box plots shown in \figref{fig_Pf_SSB}. Using such small datasets for MCS to estimate $P_f$ on the order of $10^{-3}$ is generally not recommended. However, we include these results to show that the emulators provide reasonable estimates with limited data. The bias in the emulator estimates decreases as the experimental design size increases, ultimately converging to the reference probability of failure, $P_f$. Additionally, as observed in the previous example, the box plots obtained by the emulators are consistently narrower than those from MCS, indicating that the emulators converge faster to $P_f$. These results support that emulators are a viable alternative to MCS in the context of reliability analysis on stochastic models.

\subsection{Wind turbine application}

Understanding and predicting the performance of wind turbines under realistic wind conditions is paramount when designing them. Consequently, reliability analysis of stochastic models has often focused on wind turbines, as shown in numerous studies \cite{Choe_2015,Choe_2016,Choe_2017,Cao_2019,Pan_2020,Li_2021}. The typical design process involves two steps: generating wind excitations and conducting aero-elastic simulations to model the interaction between wind inflow, aerodynamics, and structural dynamics. While the aero-elastic model itself is deterministic, stochasticity arises from the wind excitation. The wind loads are represented as 10-minute random fields defined by \textit{macroscopic parameters} such as wind speed and turbulence intensity. Although these parameters simplify wind modeling, they do not fully characterize the wind inflow. Consequently, multiple wind fields can be generated from the same parameters, requiring a stochastic treatment of the problem.

We test our methodology using the dataset from \cite{Barone_2012}, which includes the equivalent of 96 years of operational data for the onshore 5 MW NREL reference turbine \citep{Jonkman_2009_WindTurbine}. Stochastic wind inflow is generated using the NREL TurbSim code \citep{Jonkman_2009_TurbSim}, with average wind speed and turbulence intensity as inputs. The wind speed is modeled using a truncated Rayleigh distribution, following \cite{Moriarty_2008}. The average wind speed $U$ is 10 m/s (untruncated), with cut-in and cut-off speeds of 3 m/s and 25 m/s, respectively. Turbulence intensity is specified deterministically as a function of mean wind speed, following the IEC normal turbulence model \citep{iec_2005_61400}, making wind speed the only random variable in this study. Aero-elastic simulations are performed using the FAST code \citep{Jonkman_2005,jonkman2013addendum}. 

Of interest is the probability that the maximum flap-wise bending moment at the root of the blade $M_b$, within the simulated time frame, exceeds a given threshold $\tau$. The associated limit-state function can be defined as follows:
\begin{equation}
g_s(U, \omega) = \tau - M_b(U, \omega),
\end{equation}

\noindent where $M_b(U, \omega)$ represents the maximum flap-wise bending moment during the 10-minute simulations.

\begin{figure}[H]
     \centering
     \begin{subfigure}[c]{0.49\textwidth}
         \centering
         \includegraphics[width=0.9\textwidth]{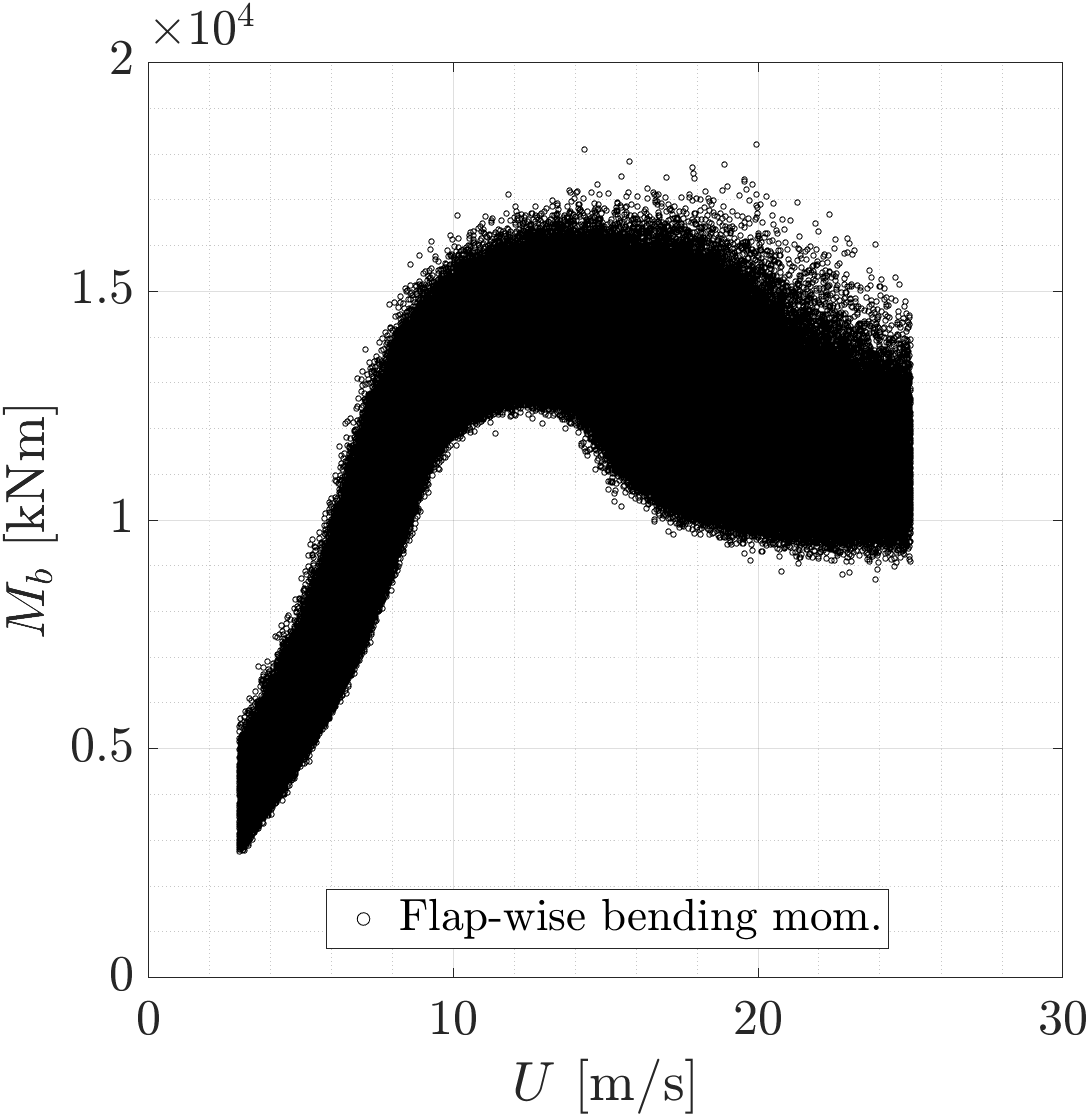}
         \caption{}
    \label{fig_dataset_bending_moment}
     \end{subfigure}
     \begin{subfigure}[c]{0.49\textwidth}
     \vspace{2mm}
         \centering
         \includegraphics[width=0.915\textwidth]{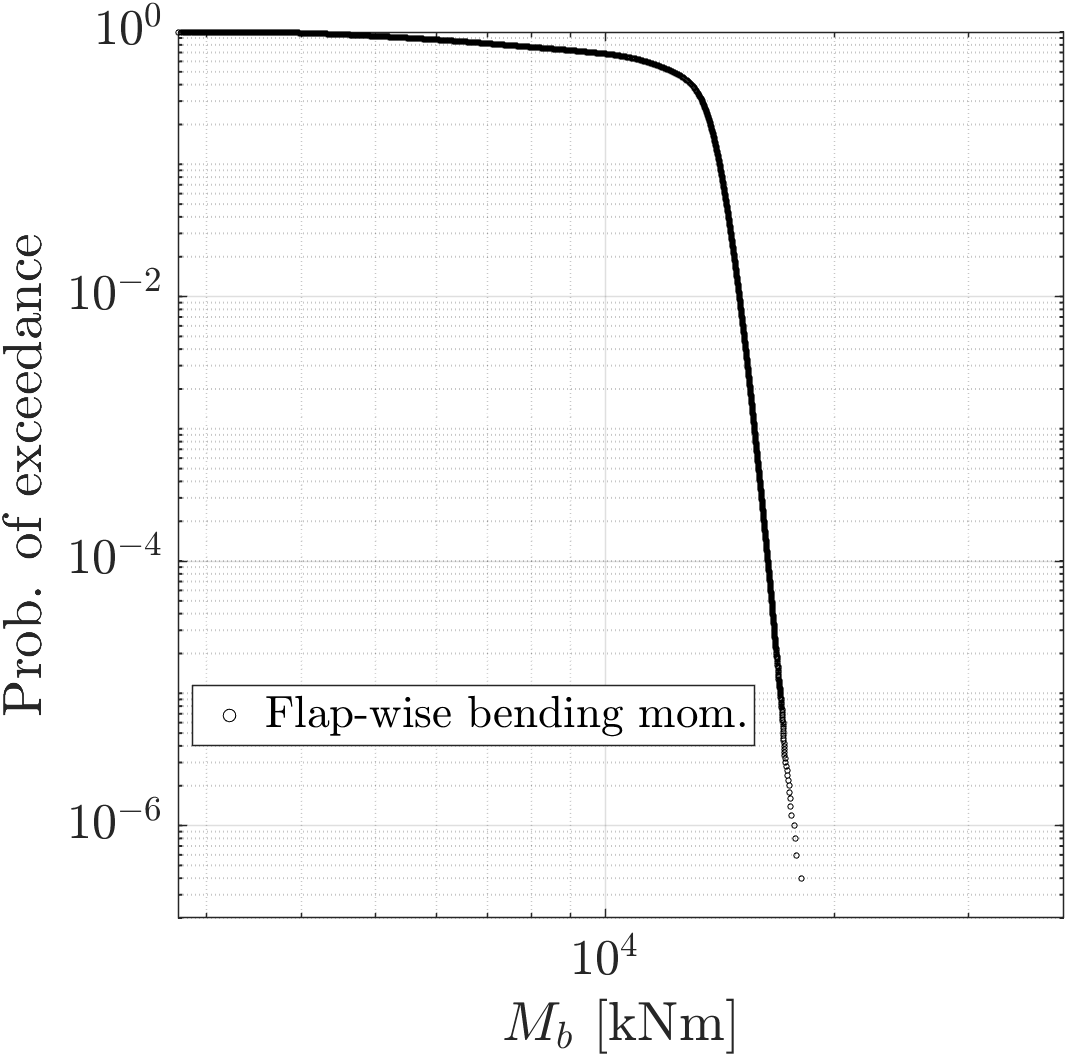}
         \caption{}
    \label{fig_dataset_survival_curve}
     \end{subfigure}
     \caption{Wind turbine application -- (a): Scatter plot depicting the relationship between the average wind speed and the maximum flap-wise bending moment for the given dataset (b): Probability of exceedance (POE) curve for the flap-wise bending moment. The x-axis represents the flap-wise bending moments (kNm), and the y-axis shows the probability that a moment on the x-axis will be exceeded.}
	 \label{fig_dataset}
\end{figure}

\figref{fig_dataset} shows the available dataset. Specifically,  \figref{fig_dataset_bending_moment} displays $M_b(U, \omega)$ as a function of the average wind speed $U$. \figref{fig_dataset_survival_curve} depicts the exceedance probability curve, which indicates the probability of obtaining a maximum moment that exceeds a threshold $\tau$ during the simulation period, given the dataset. In this example, the focus is on computing the exceedance probability curve using the emulators. These probabilities are later converted into return periods, which are used for designing.
We train the emulators on a subset of the available data, considering different experimental design sizes $N \in \acc{500, 15{,}000, 50{,}000}$. The polynomial chaos expansions used to surrogate the parameters $\lambda_{1,2}(\ve{x})$ in the GLaM model had adaptive degrees ranging from $p \in [1, 10]$, while the expansions for $\lambda_{3,4}(\ve{x})$ varied from $p \in [0, 3]$. For the SPCE emulators, the degree adaptivity ranged from $p \in [3, 12]$. Since this example involves only a single random variable, hyperbolic truncation does not apply. 50 different experiments are conducted to account for statistical variability. 

Since this is a 1D example, we can visually compare the results obtained from the emulators with those from the dataset. In our results, we show the comparison for the mean function $\mu(u)$ and $\alpha$-quantile function $Q_{\alpha}(u)$. Since the data is discrete, we estimate these functions using a moving window approach. For each wind speed $u$, we define a window $[u - \Delta, u + \Delta]$, where $\Delta = 0.05$ is a small positive value. The estimate of the mean function at $u$ is the empirical mean of the values $M_b$ within this window:
\begin{equation}
\mu(u) = \frac{1}{N_w} \sum_{i: u^{\prt{i}} \in [u - \Delta, u + \Delta]} M_b^{\prt{i}},
\label{eq_empirical_mean}
\end{equation}

\noindent where $M_b^{\prt{i}} = M_b\prt{u^{\prt{i}}}$ and $N_w$ is the number of data points within the window:
\begin{equation}
N_w = \left| \{i: u^{\prt{i}} \in [u - \Delta, u + \Delta]\} \right|.
\end{equation}

The empirical $\alpha$-quantile is estimated by first sorting the values $M_b^{\prt{i}}$ within the window. The $\alpha$-quantile, $Q_{\alpha}(u)$, is then defined as:
\begin{equation}
Q_{\alpha}(u) = M_b^{(\lfloor \alpha N_w \rfloor)}
\label{eq_empirical_quantile}
\end{equation}

\noindent where $\lfloor \cdot \rfloor$ represents the floor function, which gives the largest integer less than or equal to $\alpha N_w$.

\begin{figure}[H]
     \centering
     \begin{subfigure}[c]{0.49\textwidth}
         \centering
         \includegraphics[width=0.8\textwidth]{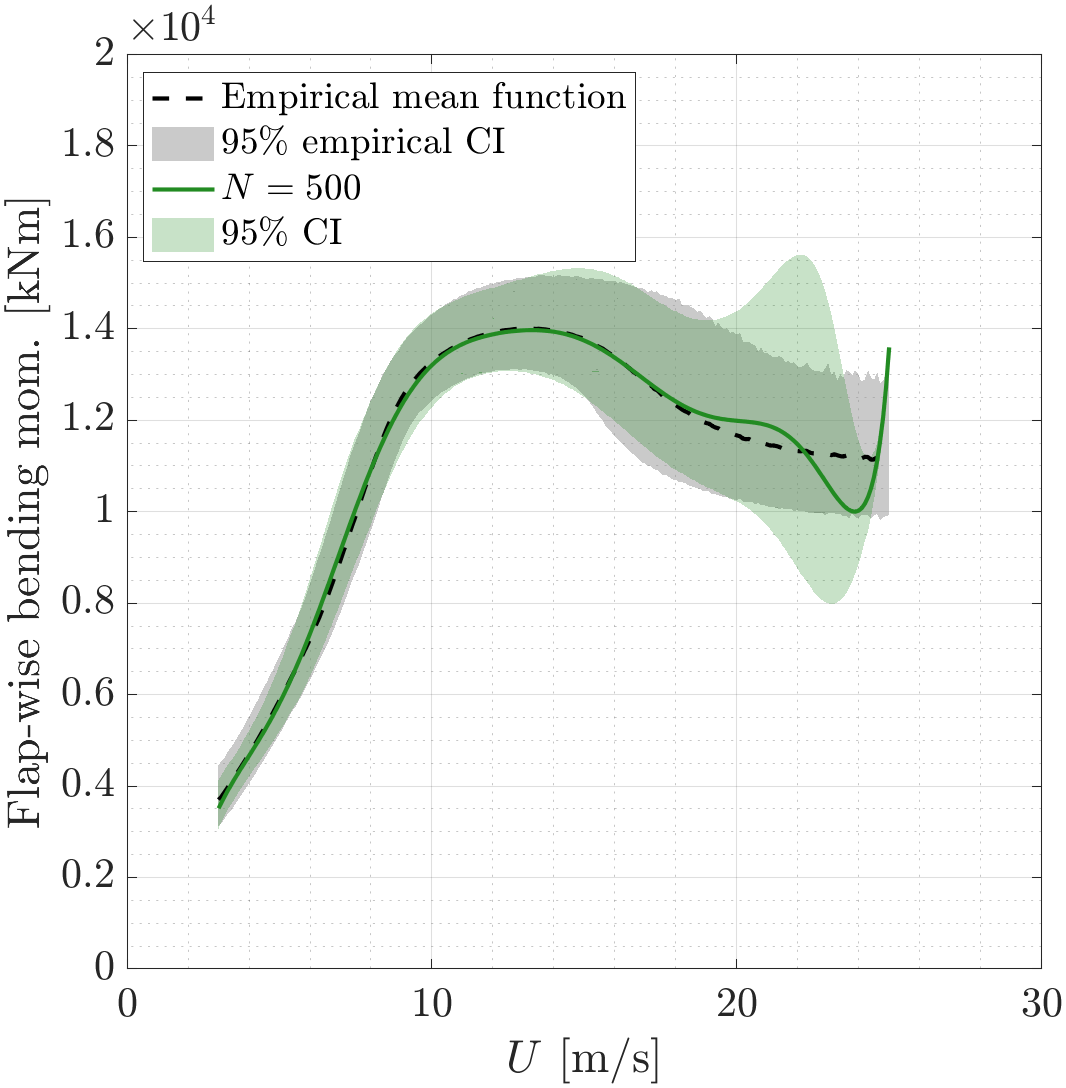}
         \caption{}
    \label{}
     \end{subfigure}
     \begin{subfigure}[c]{0.49\textwidth}
         \centering
         \includegraphics[width=0.8\textwidth]{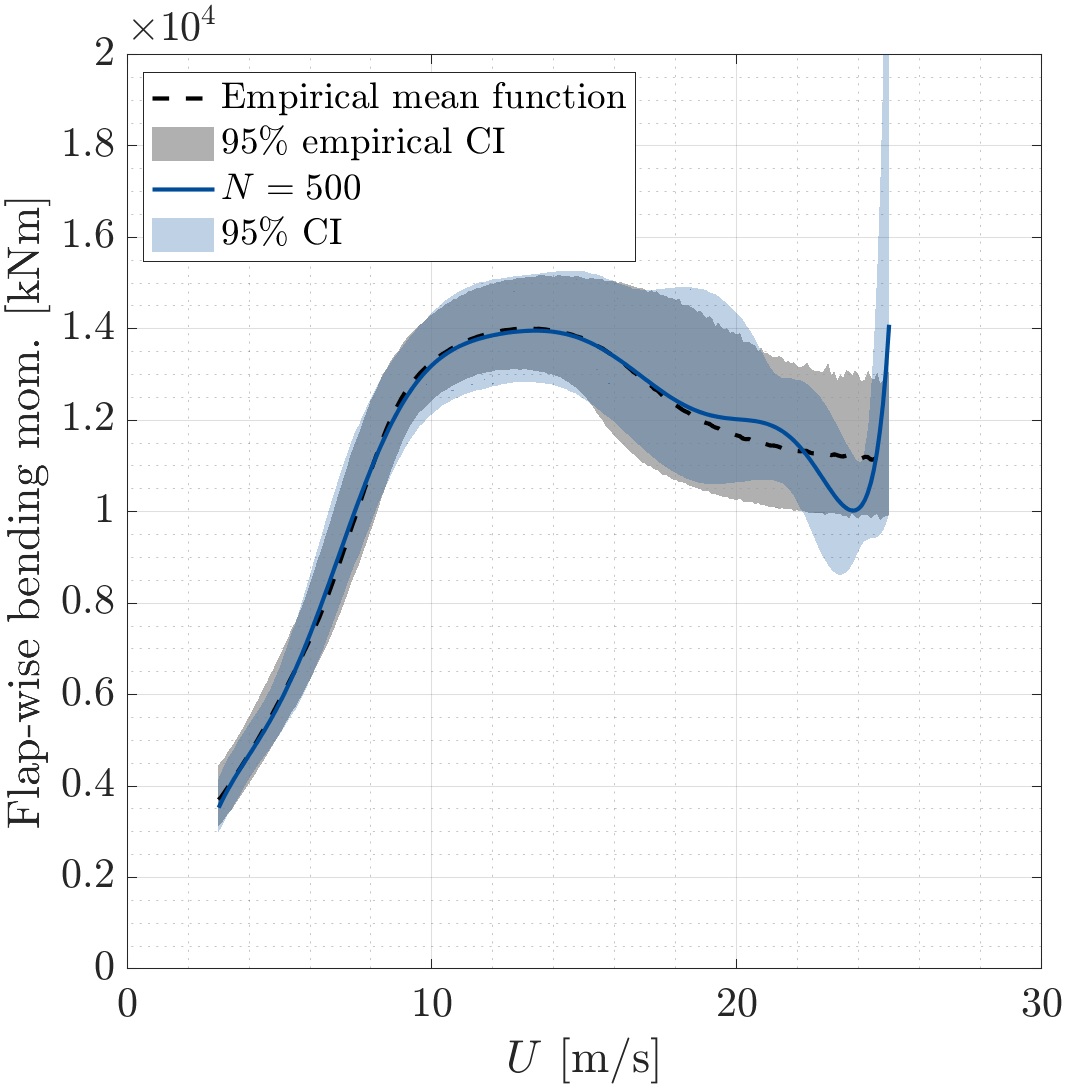}
         \caption{}
    \label{}
     \end{subfigure}
     \hspace{2mm}
          \begin{subfigure}[c]{0.49\textwidth}
         \centering
         \includegraphics[width=0.8\textwidth]{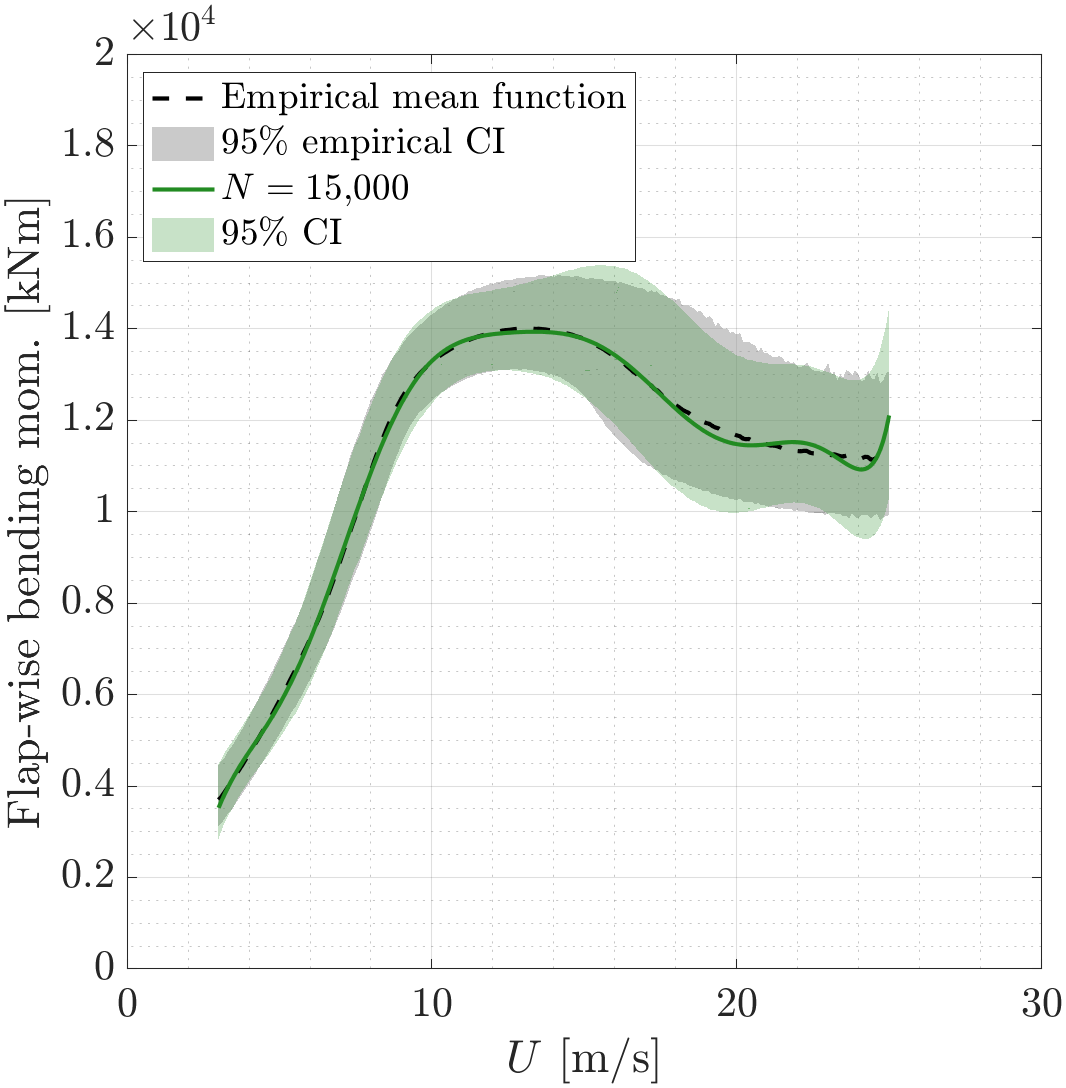}
         \caption{}
    \label{}
     \end{subfigure}
     \begin{subfigure}[c]{0.49\textwidth}
         \centering
         \includegraphics[width=0.8\textwidth]{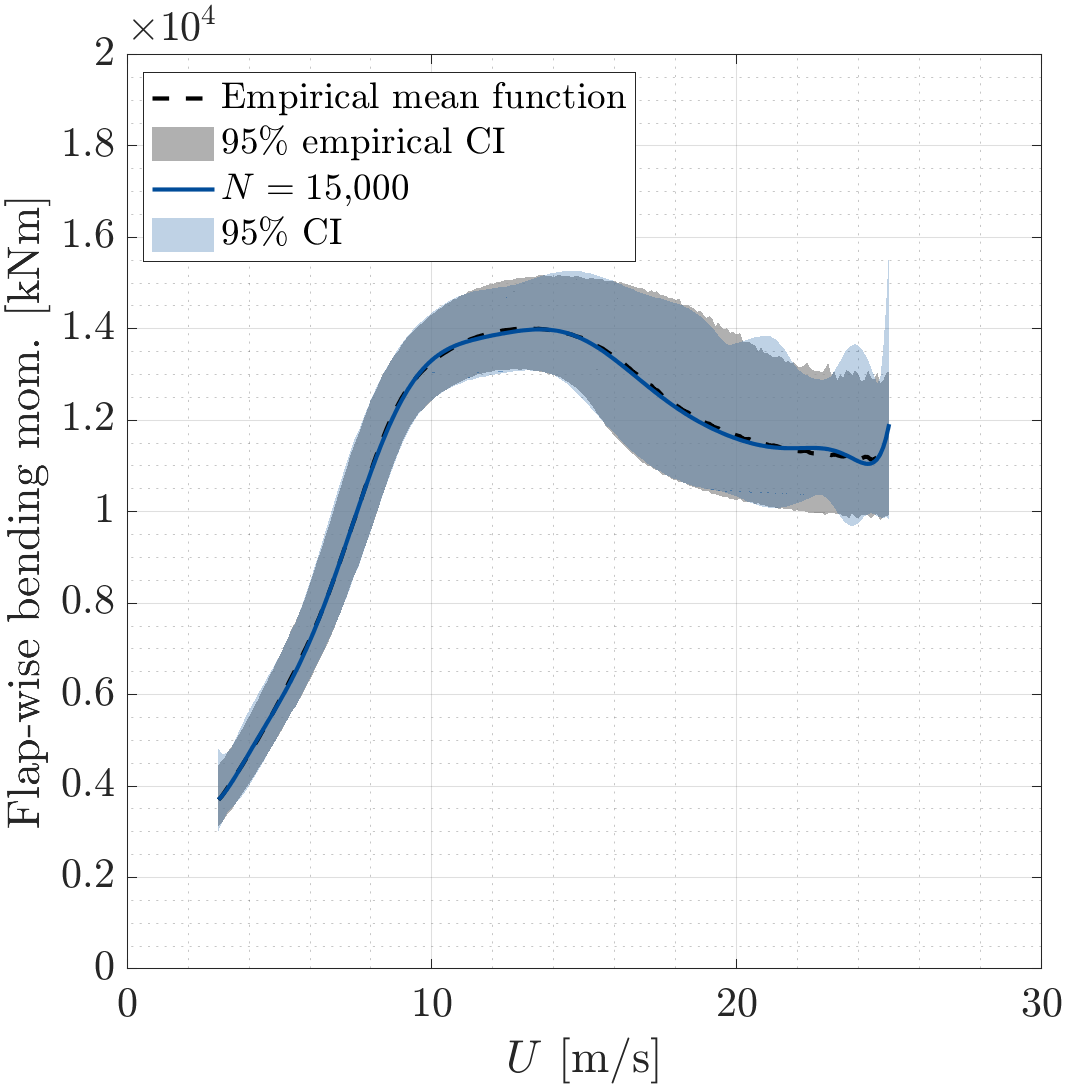}
         \caption{}
    \label{}
     \end{subfigure}
          \hspace{2mm}
          \begin{subfigure}[c]{0.49\textwidth}
         \centering
         \includegraphics[width=0.8\textwidth]{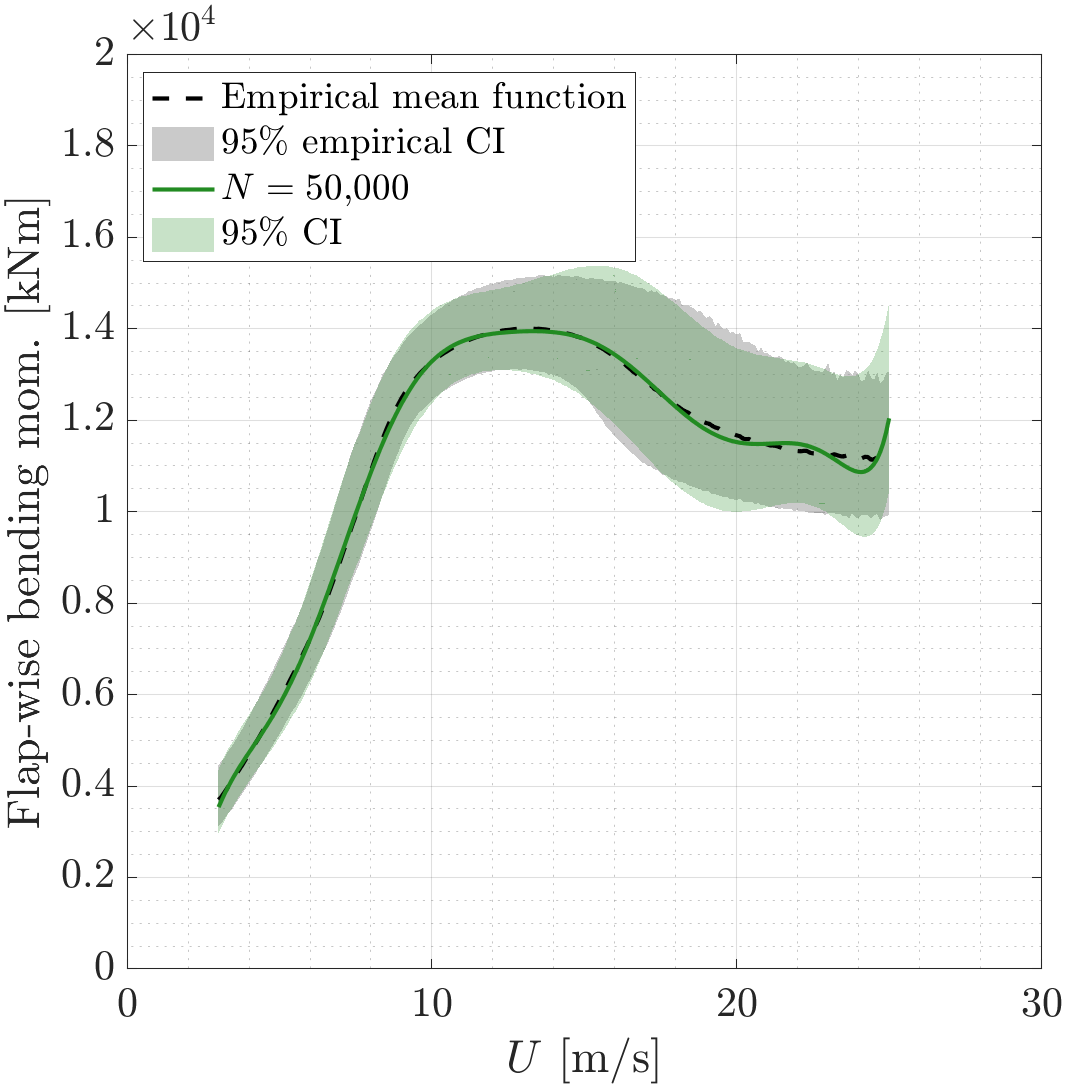}
         \caption{}
    \label{}
     \end{subfigure}
     \begin{subfigure}[c]{0.49\textwidth}
         \centering
         \includegraphics[width=0.8\textwidth]{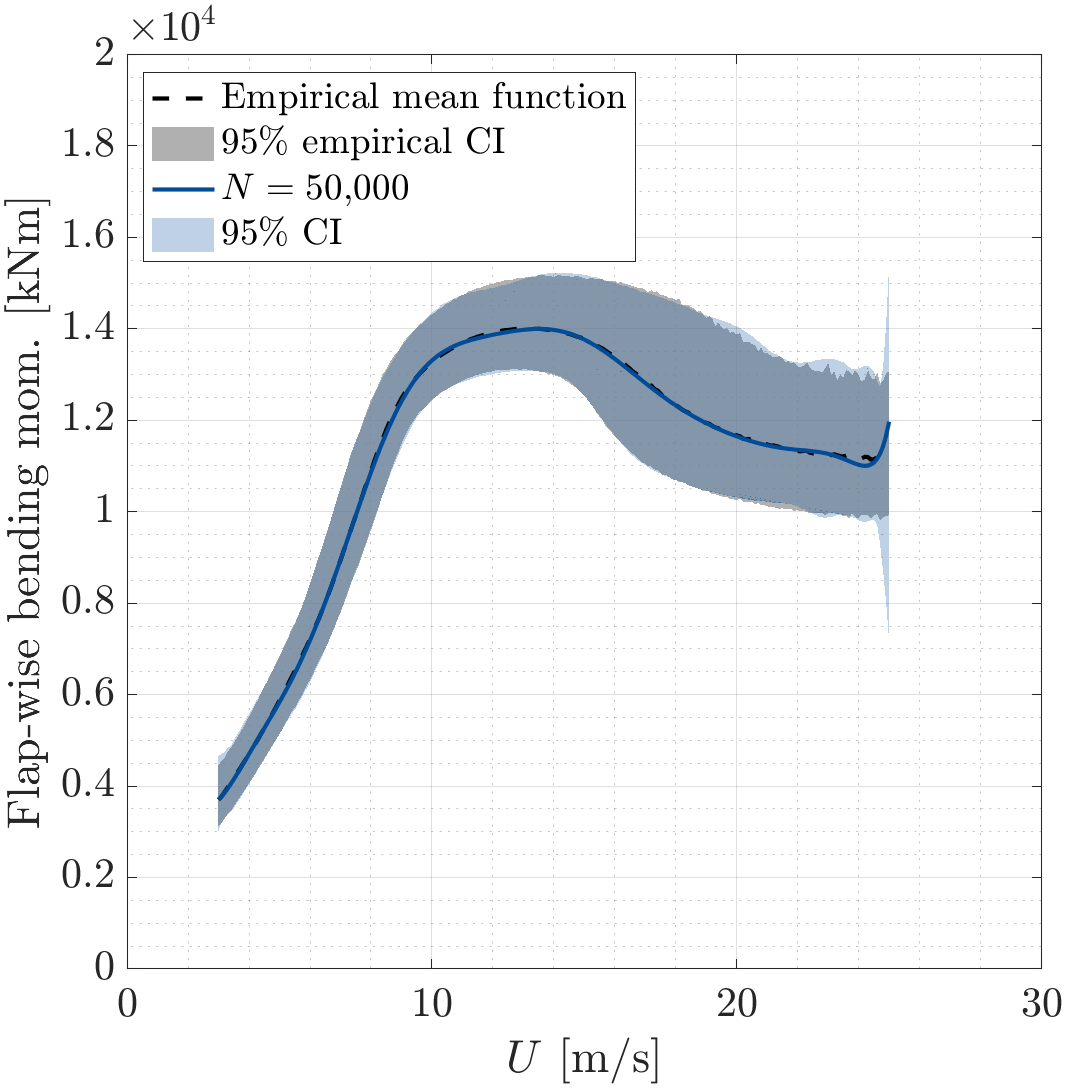}
         \caption{}
    \label{}
     \end{subfigure}
     \caption{Wind turbine application -- Plots illustrating the empirical mean curve (dotted line) and the corresponding 95\% empirical confidence interval (shaded gray area) based on the available dataset. The full, colored line represents the mean function obtained by the emulator, while the colored shaded area shows the 95\% centered confidence interval generated by the emulators when trained with an ED of size $N$.}
	 \label{fig_stochastic_emulators_fit_WindTurbine}
\end{figure}

\figref{fig_stochastic_emulators_fit_WindTurbine} shows the fitting of the emulators for different experimental design sizes. The colored lines represent the mean functions obtained by the emulators, while the dashed black lines show the empirical mean computed from the data (\eqrefe{eq_empirical_mean}). The shaded colored areas indicate the $95\%$ confidence intervals obtained by the emulators, and the grey-shaded area represents the empirical confidence interval from the simulator $Q_{97.5\%}- Q_{2.5\%}$ (\eqrefe{eq_empirical_quantile}). 

With a relatively small experimental design of 500 points, the surrogate model captures the behavior of the mean function fairly accurately. Indeed, noticeable discrepancies occur at high wind speeds. These discrepancies are due to the limited number of training points in the tails, which makes fitting in those areas more challenging. However, as the experimental design size increases, the accuracy of the emulators improves, and the difference in the tail reduces. For large designs, such as $N=50{,}000$ points, the emulators provide more robust estimates across the domain.

\begin{figure}[H]
     \centering
     \begin{subfigure}[c]{0.49\textwidth}
         \centering
         \includegraphics[width=0.8\textwidth]{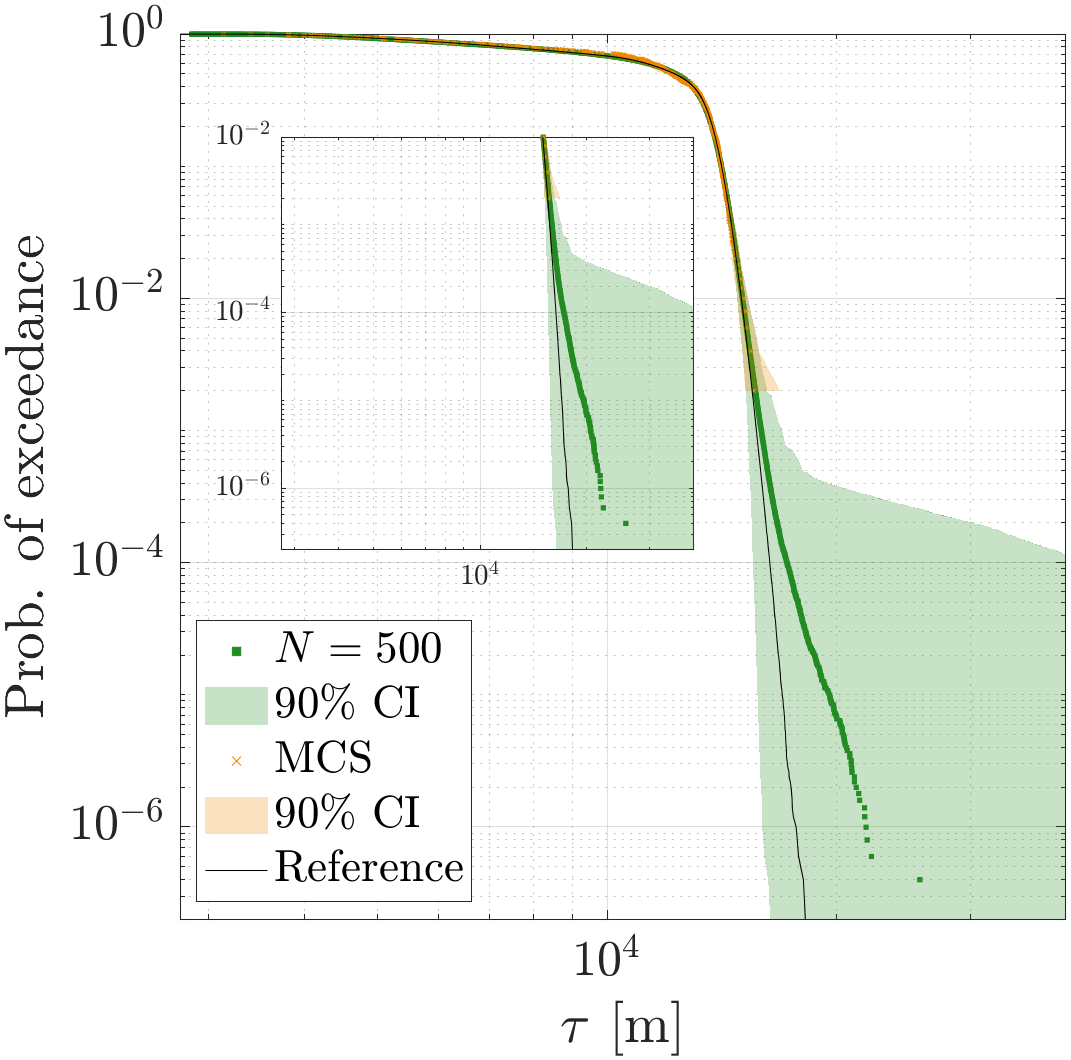}
         \caption{}
    \label{fig_WT_GLaM_ED1}
     \end{subfigure}
     \begin{subfigure}[c]{0.49\textwidth}
         \centering
         \includegraphics[width=0.8\textwidth]{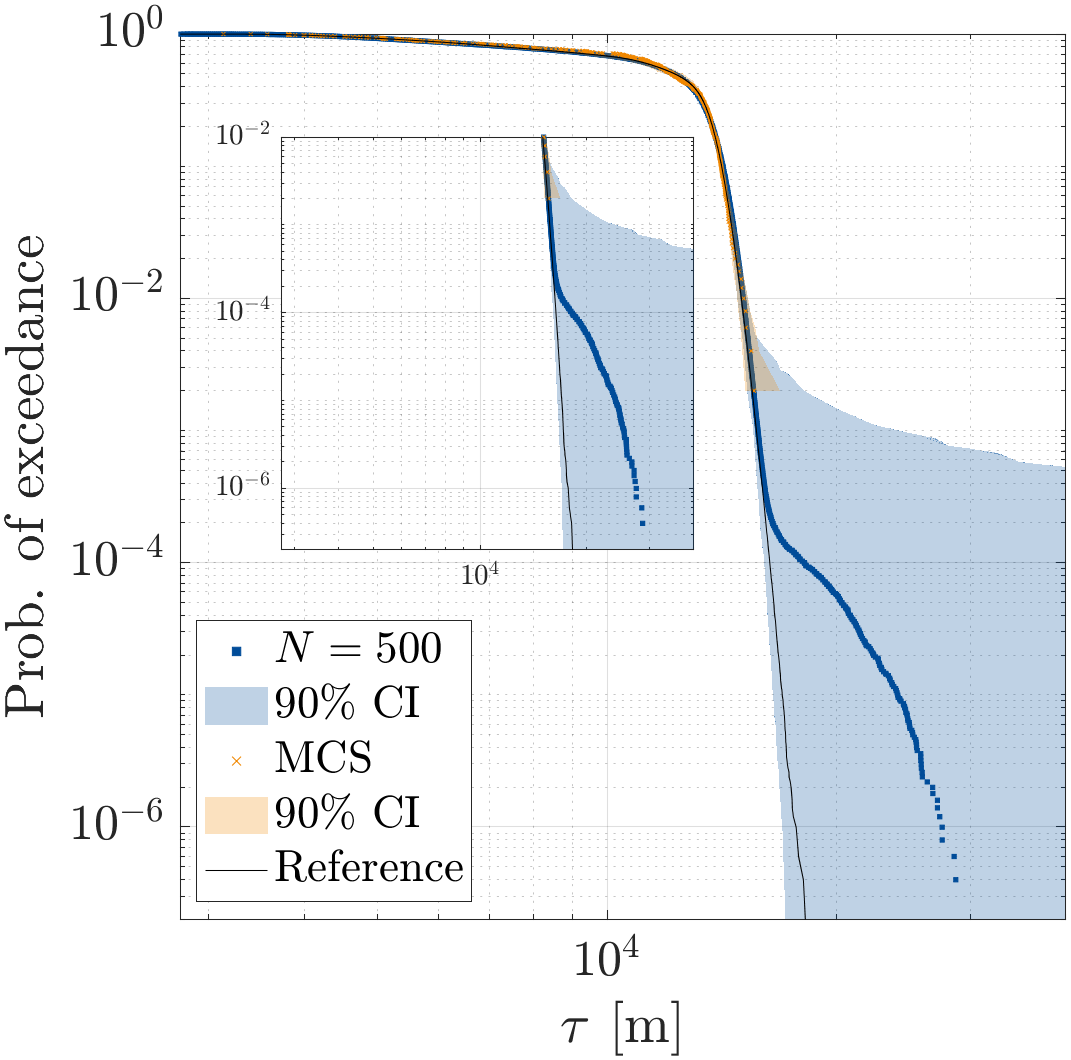}
         \caption{}
    \label{fig_WT_SPCE_ED1}
     \end{subfigure}
     \hspace{2mm}
          \begin{subfigure}[c]{0.49\textwidth}
         \centering
         \includegraphics[width=0.8\textwidth]{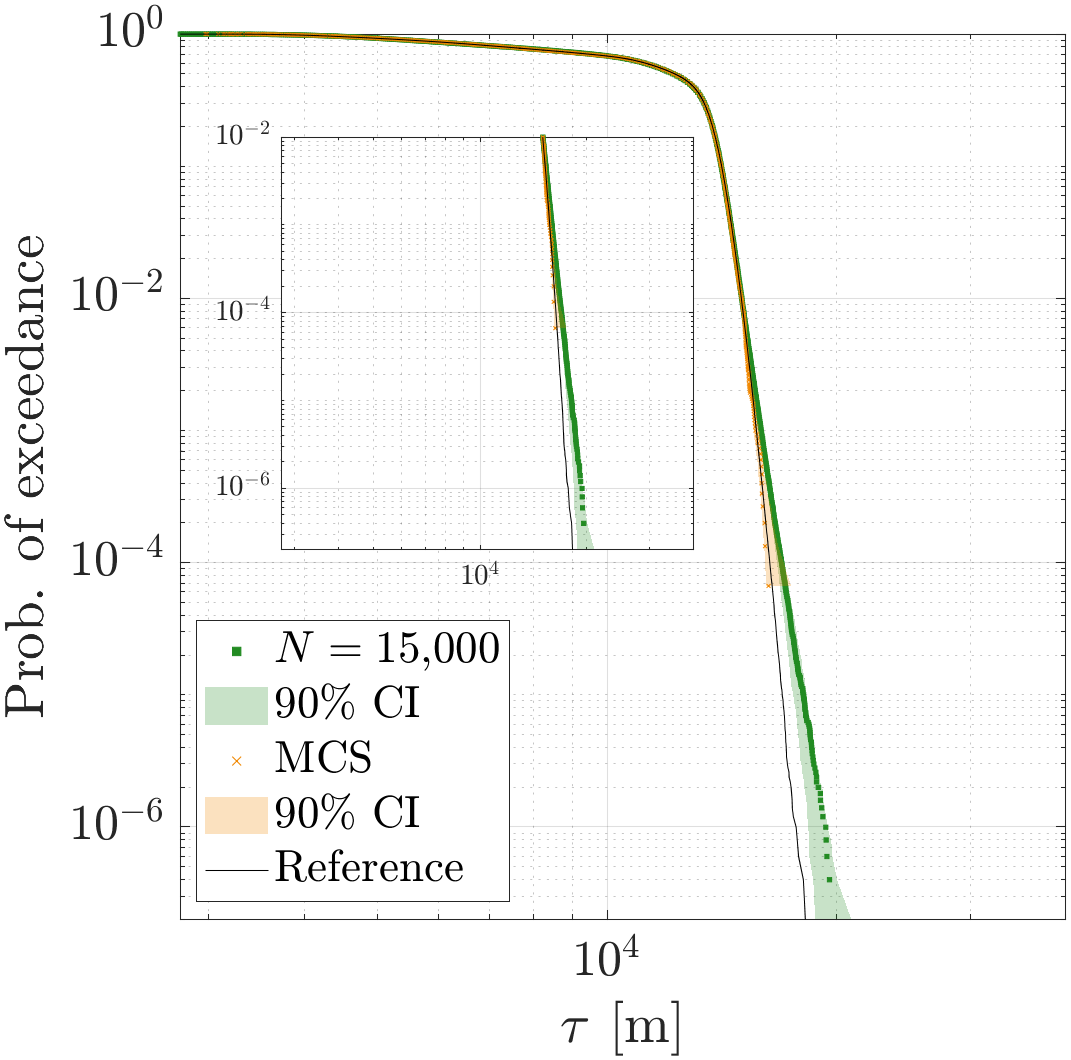}
         \caption{}
    \label{fig_WT_GLaM_ED2}
     \end{subfigure}
     \begin{subfigure}[c]{0.49\textwidth}
         \centering
         \includegraphics[width=0.8\textwidth]{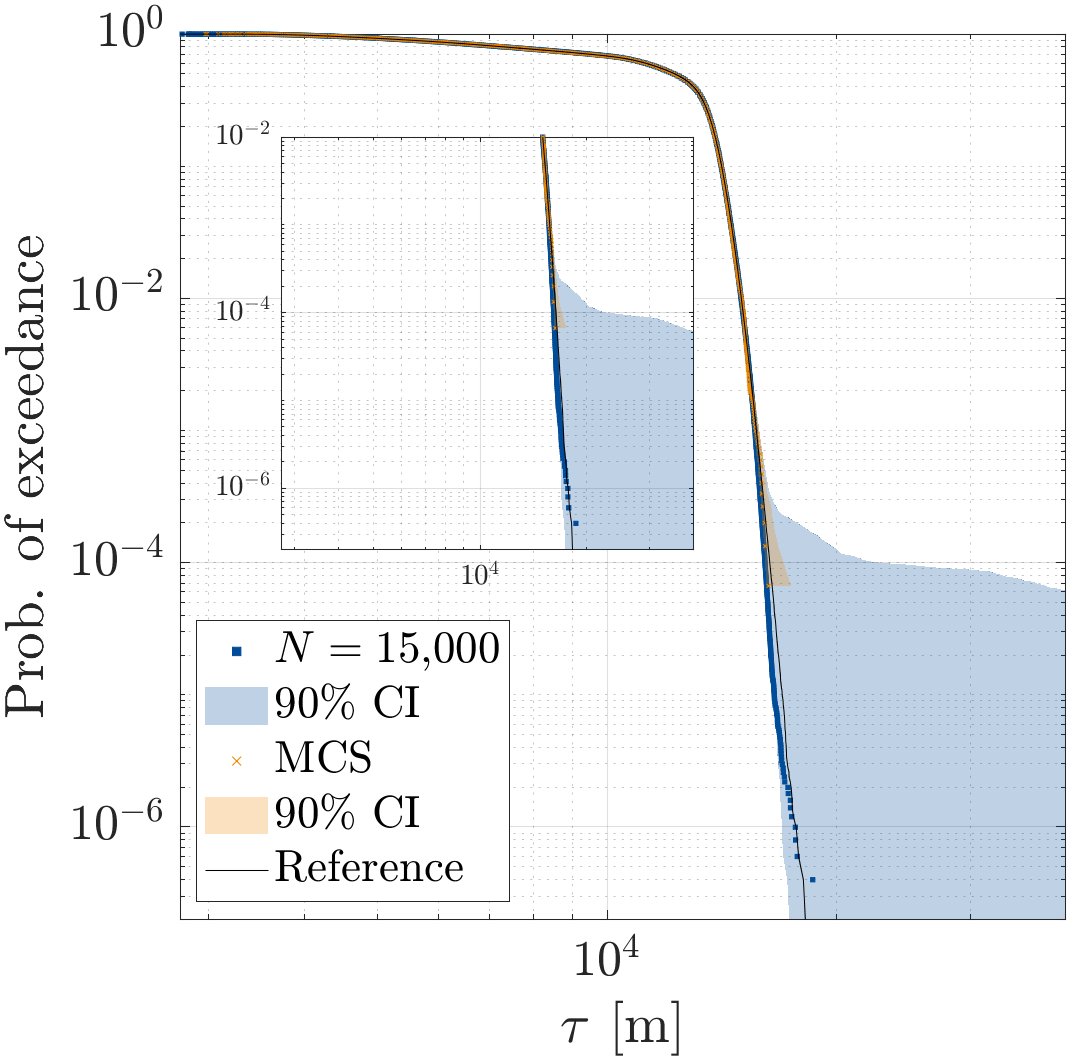}
         \caption{}
    \label{fig_WT_SPCE_ED2}
     \end{subfigure}
          \hspace{2mm}
          \begin{subfigure}[c]{0.49\textwidth}
         \centering
         \includegraphics[width=0.8\textwidth]{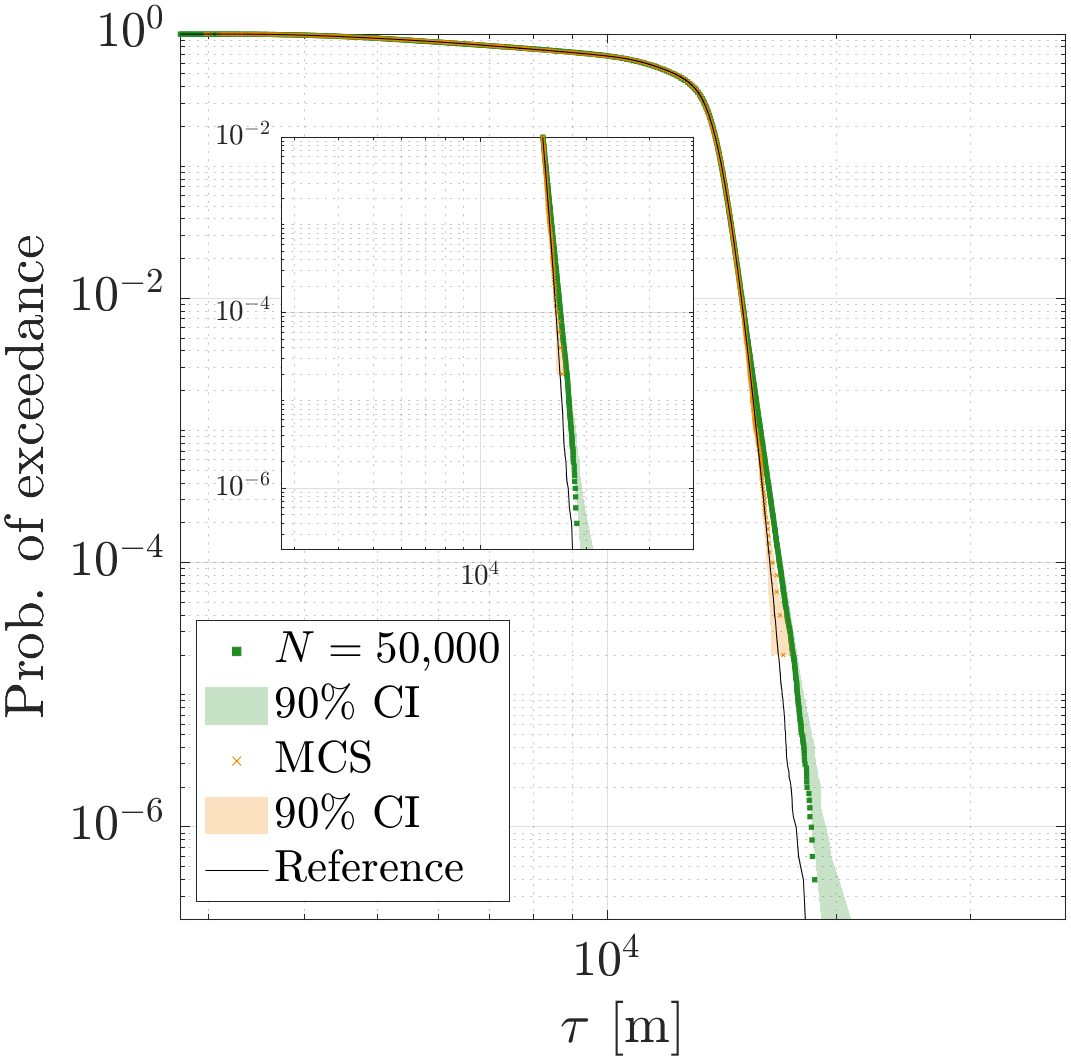}
         \caption{}
    \label{fig_WT_GLaM_ED3}
     \end{subfigure}
     \begin{subfigure}[c]{0.49\textwidth}
         \centering
         \includegraphics[width=0.8\textwidth]{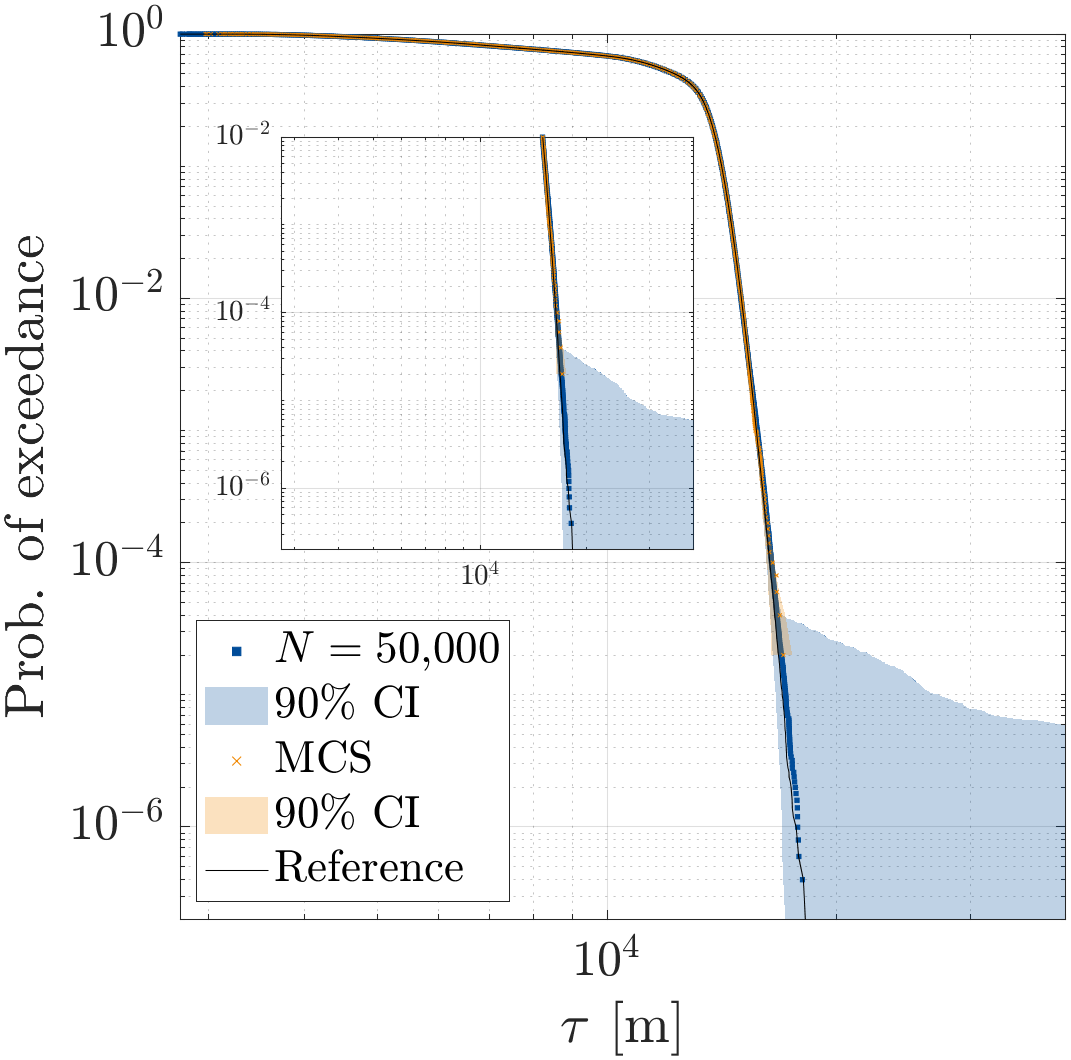}
         \caption{}
    \label{fig_WT_SPCE_ED3}
     \end{subfigure}
     \caption{Wind turbine application -- Plots illustrating the empirical mean curve (dotted line) and the corresponding 95\% empirical confidence interval (shaded gray area) based on the available dataset. The full, colored line represents the mean function obtained by the emulator, while the colored shaded area shows the 95\% centered confidence interval generated by the emulators when trained with an ED of size $N$.}
	 \label{fig_POE}
\end{figure}

\figref{fig_POE} shows the exceedance probability curve obtained from the dataset (black continuous line) and those obtained from the emulators (green for GLaM and blue for SPCE) for different experimental designs. The results of the direct Monte Carlo simulation are shown in orange. Shaded regions around each curve represent the 90\% central confidence intervals based on 50 repetitions of the simulations, visually presenting the variance observed across experiments. The results depicted in \figref{fig_POE} indicate that the emulators generally capture the behavior of the exceedance curve, even for relatively small experimental design sizes, as shown in \figref{fig_WT_GLaM_ED1,fig_WT_SPCE_ED1}. As the experimental design size increases, the curves approach the reference one.

While MCS is limited to the available data, the emulators allow for some degree of extrapolation. As expected, the accuracy decreases in the extrapolation region, which is reflected in the sudden increase in the shaded region. Nevertheless, the emulators continue to provide estimates beyond the data range. Regarding the variance comparison, GLaM consistently produces smaller variances than MCS, where data is available. On the other hand, SPCE exhibits higher variances due to a model misfit for wind speeds $U \geq 25$ m/s as shown in \figref{fig_stochastic_emulators_fit_WindTurbine}. This misfit causes SPCE to predict much higher moments than those observed in the original simulator, resulting in an increase in the shaded area. This drawback can be addressed through active learning techniques. In this case, the strategy would identify that the tail is poorly fitted and add points in that region to improve the predictions of the emulator.

\section{Conclusion}

In this paper, we investigated structural reliability analysis for problems involving stochastic simulators. The inherent randomness in these models impacts system reliability and needs to be considered in the analysis. As a result, the probability of failure can be defined in different ways. We show that calculating the probability of failure using MCS is possible, but similar to the deterministic case, it remains computationally expensive. To overcome this, we propose using GLaM and SPCE as surrogate models for stochastic simulators in the context of reliability analysis. The main advantage of these emulators is that they do not require replicated data for training. This allows us to use existing data to improve preliminary estimates and gain better insights into the system. We demonstrated that when the conditional probability of failure function is known, it provides a lower variance estimator compared to direct MCS. We also show how this function can be obtained semi-analytically from the emulators.

Our results show that these emulators provide accurate estimates of the reference probability of failure. Additionally, they consistently reduce the variance of the estimates, meaning they can be seen as a variance-reduction technique, allowing faster convergence when compared to MCS. The main limitation of our approach is the need for a relatively large experimental design to achieve accuracy. This is due to the global training method we used for the emulators. However, in reliability analysis, global accuracy is not always necessary to get a reliable estimate of the probability of failure. Active learning approaches, which focus on refining the experimental design in important regions, could help reduce the amount of training data needed and speed up convergence.
\appendix
\section{Variance of estimator for unknown $s\prt{\ve{x}}$}\label{appendix}
\label{annex_A}

This section demonstrates that when the conditional probability of failure function $s(\ve{x})$ is unknown, estimating $P_f$ is more efficient by computing $\overline{P_f}$ rather than using a double-loop estimator, denoted as $\dbar{P_f}$. The double-loop approach becomes computationally expensive because of the inner MCS loop to estimate $s(\ve{x})$ for each $\ve{x}$. The double-loop estimator can be written as:
\begin{equation}
\begin{split}
\dbar{P_f} &= \frac{1}{N} \sum_{i=1}^{N} s\prt{\ve{x}^{(i)}},\\
&= \frac{1}{N} \sum_{i=1}^{N} \frac{1}{M} \sum_{j=1}^{M} \I\prt{\ve{x}^{(i)},\ve{z}^{(i,j)}},\\
\end{split}
\end{equation}

where $N$ is the number of samples of $s(\ve{x})$ evaluated, and $M$ is the number of samples used in MCS to compute each $s(\ve{x})$.

To show that $\overline{P_f}$ converges faster than $\dbar{P_f}$, we compare their variances for a given budget. Using a notation consistent with Sec.~\ref{sec_reliability_on_emulators}, we define the total budget as $N_{\text{MCS}} = N M$. The variance of the double-loop estimator $\dbar{P_f}$ is:
\begin{equation}
    \begin{split}
        \Var{\dbar{P_f}} &= \Var{\frac{1}{N} \sum_{i=1}^{N} \frac{1}{M} \sum_{j=1}^{M} \I\prt{\ve{x}_i,\ve{z}_{i}^{j}}}\\
        &=\frac{1}{N^2} \sum^N_{i=1}\frac{1}{M^2} \sum_{j=1}^M \sum_{k=1}^M \Cov{\I\prt{\ve{x}_i, \ve{z}_i^j}, \I\prt{\ve{x}_i, \ve{z}_i^k}}
    \end{split}
\end{equation}

Assuming that $\ve{X}$ and $\ve{Z}$ are independent and identically distributed (i.i.d.), the expression simplifies to:
\begin{equation}
    \begin{split}
        \Var{\dbar{P_f}} &= \frac{1}{N^2} \sum^N_{i=1}\frac{1}{M^2} \sum_{j=1}^M \prt{\Var{\I\prt{\ve{X}, \ve{Z}}} + \sum_{j\neq k} \Cov{\I\prt{\ve{X},\ve{Z}^j}, \I\prt{\ve{X},\ve{Z}^k}}}\\
        &=  \frac{1}{N^2} N\frac{1}{M^2} M\prt{\Var{\I\prt{\ve{X}, \ve{Z}}} + \prt{M-1}\Var{s\prt{\ve{X}}}}\\ 
        &=  \frac{1}{N}\frac{1}{M} \prt{\Var{\I\prt{\ve{X}, \ve{Z}}} + \prt{M-1}\Var{s\prt{\ve{X}}}}\\
        &=\frac{1}{N_{\text{MCS}}}\prt{\Var{\I\prt{\ve{X}, \ve{Z}}} + \prt{M-1}\Var{s\prt{\ve{X}}}}
    \end{split}
    \label{eq_var_double_loop}
\end{equation}

Thus, for a fixed computational budget $N_{\text{MCS}}$, the variance of $\dbar{P_f}$ is larger than the variance of $\overline{P_f}$  (\eqrefe{var_single_loop}) due to the additional term $(M-1)\Var{s(\ve{X})}$ in the expression. This shows that estimating $\overline{P_f}$ is more efficient than $\dbar{P_f}$ when $s(\ve{x})$ is unknown.

\section*{Acknowledgments}
The support of European Union’s Horizon 2020 research and innovation program under the Marie Skłodowska-Curie grant agreement No 955393 is greatly acknowledged.

Add acknowledgment to Nora for the appendix.

\newpage
\bibliography{References} 
\end{document}